\documentclass[aps,amsmath,amssymb,prb,twocolumn,showpacs,superscriptaddress]{revtex4}
\usepackage{longtable}
\usepackage{bm}
\usepackage{epsfig}
\usepackage[dvips]{color}

\newcommand{\nn}{\nonumber}

\newcommand{\tav}[1]{\left\langle #1\right\rangle} 
\newcommand{\cav}[1]{\left\langle\left\langle #1\right\rangle\right\rangle} 
\newcommand{\dav}[1]{\overline{#1}}                

\newcommand{\gb}[1]{\left[#1\right]}

\newcommand{\Tr}{{\operatorname{Tr}}}
\newcommand{\Ei}{{\operatorname{Ei}}}

\renewcommand{\vec}[1]{{\bf #1}}

\begin{document}

\title{Influence of thermal fluctuations on quantum phase transitions in 
one-dimensional disordered systems: Charge density waves and
Luttinger liquids}

\author{Andreas Glatz}
\affiliation{Institut f\"ur Theoretische Physik, Universit\"at zu
K\"oln, Z\"ulpicher Str. 77, 50937 K\"oln, Germany}
\affiliation{Materials Science Division, Argonne National
Laboratory, Argonne, Illinois 60439}
\author{Thomas Nattermann}
\affiliation{Institut f\"ur Theoretische Physik, Universit\"at zu
K\"oln, Z\"ulpicher Str. 77, 50937 K\"oln, Germany}

\date{\today}

\begin{abstract}
The low temperature phase diagram of 1D weakly disordered quantum
systems like charge or spin density waves and Luttinger liquids is
studied by a \emph{full finite temperature} renormalization group
(RG) calculation. For vanishing quantum fluctuations this approach
is amended by an \emph{exact} solution in the case of strong
disorder and by a mapping onto the \emph{Burgers equation with
noise} in the case of weak disorder, respectively. At \emph{zero}
temperature we reproduce the quantum phase transition between a
pinned (localized)  and an unpinned (delocalized) phase for weak
and strong quantum fluctuations, respectively, as found previously
by Fukuyama or Giamarchi and Schulz.

At \emph{finite} temperatures the localization transition is
suppressed: the random potential is wiped out by thermal
fluctuations on length scales larger than the thermal de Broglie
wave length of the phason excitations. The existence of a zero
temperature transition is reflected in a rich cross-over phase
diagram of the  correlation functions. In particular we find four
different scaling regions: a \emph{ classical disordered}, a
\emph{quantum disordered}, a \emph{quantum critical} and a
\emph{thermal} region. The results can be transferred  directly to
the discussion of the influence of disorder in superfluids.
Finally we extend the  RG calculation to the treatment of a
commensurate lattice potential. Applications to related  systems
are discussed as well.
\end{abstract}

\pacs{ 71.10.Pm, 72.15.Rn/Nj, 73.20.Mp/Jc }

\maketitle

\section{Introduction}

The collective behavior of condensed modulated structures like
charge or spin density waves
(CDWs/SDWs)~\cite{Gruener,Gruener94,Brazovski99}, flux line
lattices~\cite{Blatter,NatSchei} and Wigner
crystals~\cite{Brazovski99} in random environments has been the
subject of detailed investigations since the early 1970s. These
were motivated by the drastic influence of disorder: without
pinning CDWs would be ideal superconductors whereas type-II
superconductors would show finite resistivity.  In three
dimensional systems the low temperature phase of these structures
is determined by a zero temperature disorder fixed point resulting
in quasi-long-range order and glassy dynamics (for recent reviews
and further references see [\onlinecite{Blatter,NatSchei}]). In
two dimensions this fixed point is extended to a fixed line which
terminates at the glass transition
temperature~\cite{CardyOst,ViFer84}.  In the low temperature
phase, correlations of the positional order decay slightly faster
than a power law and the linear resistivity
vanishes~\cite{NatSchei}.

In one dimension the situation is different: the glass temperature
is shifted to $T=0$. Nevertheless, there remains a residual trace
of disorder which is reflected in the low temperature behavior of
spatial correlations and the dynamics~\cite{Feigel80, GlatzLi02}.
Clearly, at low temperatures also quantum fluctuations have to be
taken into account. Disorder and quantum fluctuations in $1$D CDWs
at zero temperature have been considered previously (see, e.g.,
[\onlinecite{Fukuyama84, GiaSchulz87}]) and an unpinning
(delocalization) transition as a function of the strength of
quantum fluctuations was found. Finite temperature effects were
partially incorporated by truncating the renormalization group
(RG) flow at the de Broglie wave length of the phason
excitations~\cite{GiaSchulz87}. However, for a complete study of
the thermal to quantum crossover, quantum and thermal fluctuations
have to be considered on an equal footing~\cite{Chakra}, which is
the main aim of this paper.

Experimentally, quasi--1D behavior can be seen in real materials,
e.g., in whiskers~\cite{Brock+93}, hairlike single crystal fibers
like $\text{NbSe}_3$, with a transverse extension smaller than the
correlation length or in chain like crystals with weak interchain
coupling. In the latter case there is a large crossover length
scale up to which 1D behavior can be
observed~\cite{Gruener,Brazovski99}.

The results obtained for the CDWs or SDWs have a large number of
further applications on disordered quantum systems: they relate,
e.g., to the localization transition of Luttinger
liquids~\cite{Fukuyama84,GiaSchulz87},
superfluids~\cite{FisherGrinstein}, tunnel junction
chains~\cite{Korshunov}, Josephson coupled chains of these
systems, if the coupling is treated in mean-field
theory~\cite{Fukuyama84}, and CDWs in a lattice potential.
However, we will use the terminology of CDWs in most parts of this
paper.

The remaining part of the  paper is organized as follows: In
section \ref{sec.model} we give a detailed introduction to our
model and the notation used in this paper. We also briefly discuss
the influence of Coulomb interaction on the properties of the
system. In section \ref{sec.disorder} the influence of the
disorder  is studied in detail. Using an anisotropic momentum
shell renormalization group calculation, in which the full
Matsubara sum over frequncies is performed, we obtain flow
equations for the effective strength of the disorder,   thermal
and  quantum fluctuations (i.e., the interaction strength in the
case of Luttinger liquids). These are discussed first in the case
of zero temperature and agreement with previous results is
obtained~\cite{Fukuyama84, GiaSchulz87}. At finite temperatures
the disorder always renormalizes to zero. In  the classical limit
two more methods are applied: (i) at low temperatures and strong
disorder the  ground state of the model is calculated exactly.
(ii) For weak disorder and strong thermal fluctuations a second RG
calculation,  based on the mapping onto the Burgers equation with
noise, is applied. Using all these findings, the phase diagram of
the density-density correlation function is studied in section
\ref{sec.cf}. The main result of this section is the calculation
of the low temperature quantum crossover diagram for
one-dimensional CDWs. In the following section \ref{sec.sf} we
discuss briefly the application of the results to superfluids by
using the mapping to CDWs. Some of these results were previously
presented in Ref. \onlinecite{GlaNa}.

The influence of a commensurate lattice potential on a free
density wave is considered in section \ref{sec.lp}. The full
finite temperature renormalization group flow equation for this
sine-Gordon type model are derived and resulting phase diagram is
discussed.

In the appendices we present the calculation of the
renormalization group flow equations and the derivation of the
correlation function in the strong and weak pinning limit in some
detail for the interested reader. In the final section of the
appendix we list all symbols used in this paper with corresponding
references in the paper.

\section{Model}\label{sec.model}

\subsection{charge and spin density}

In this section we  derive the effective Hamiltonian which will be
the starting point for our further treatment. The strategy of the
calculation is therefore  separated into two steps. In the first
step the system is treated in a mean-field  (MF) type
approximation applied to a microscopic Hamiltonian. This leaves us
with a slowly varying complex order parameter field for which we
derive an effective Hamiltonian. The second step involves the
consideration of the fluctuations of this order parameter, which
is the topic of this paper.

We briefly summarize now the result of the mean-field calculation:
Well below the mean-field condensation temperature
$T_{\textit{MF}}$ of the CDW, the underlying lattice will be
periodically distorted with a period $\lambda$ which is related to
the Fermi wave vector $k_F$ by $\lambda=\pi/k_F$. This distortion
of the lattice leads to the formation of a gap in the dispersion
relation at $k=\pm k_F$ which is (in one dimension) proportional
to the amplitude of the lattice modulation. For small
displacements  (which are typically smaller than $1\%$ of the
interatomic spacing~\cite{Thorne96}), the increase of the elastic
energy is smaller than the gain of electronic energy due to the
formation of the gap and hence an instability is favored. The
period of the CDW depends on the band filling factor (via
$k_F=\pi/\lambda$) and is in general at arbitrary band filling
incommensurate with the undistorted lattice (with lattice constant
$a$).

In (quasi-)one--dimensional systems~\cite{Gruener94} also SDWs can
be found, but in contrast to CDWs they arise due to
electron--electron and not to electron--phonon interaction. A SDW
can be considered to consist of two CDWs, one for spin--up  and
another for spin--down electrons (see, e.g., Fig. 5 in
[\onlinecite{Gruener94}]). Therefore the spatial modulation of
SDWs is characterized by a wave vector $Q=2k_F$, as for CDWs.

The charge or spin density $\rho(x,t)$ can be written in the
form~\cite{Gruener94,Eckern86}
\begin{equation}\label{eq.cdwdensity}
\rho(x,t)=(1+Q^{-1}\partial_x\varphi(x,t))[\rho_0+
\rho_1\cos(p\varphi(x,t)+pQx)]
\end{equation}
where  $\rho_0=Q f(T)/\pi$ and $\rho_1=2|\Delta|/(\pi g v_F)$. $g$
is the dimensionless electron-phonon coupling constant  and  $v_F$
the Fermi velocity. $\rho_1$ is proportional to $|\Delta|$, the
CDW gap or the amplitude of the complex (mean field) order
parameter
\begin{equation}\label{eq.orderparameter}
\Delta(x,t)=|\Delta(x,t)|e^{\imath\varphi(x,t)}.
\end{equation}
$f$ is the condensate density ($Y=1-f$ is the so-called {\it
Yoshida function}) related to $|\Delta|$ by
\begin{equation}
f(T)=\frac{\pi T
}{\hbar}\sum_{\omega_n}\frac{(|\Delta|/\hbar)^2}{\left(\omega_n^2+
(|\Delta|/\hbar)^2\right)^{3/2}}, \,\,\,\, \omega_n=2\pi nT/\hbar.
\end{equation}
The condensate density approaches $1$ for $T\rightarrow 0$ and
$f(T)\simeq 2(1-{T}/{T_c^{\textit{MF}}})\sim |\Delta|^2$ for $T\to
T_c^{\textit{MF}}$ . $T_c^{\textit{MF}}$ denotes the mean-field
transition temperature. For quasi one--dimensional systems
$\rho_1$ has an additional factor $\zeta^{-2}$ (the area
perpendicular to the chain).

Note, that (\ref{eq.cdwdensity}) is correct for arbitrary band
filling and, to be more precisely, is the particle density of the
charge or spin carrying particles. The particle current density
$j$ follows from (\ref{eq.orderparameter}) as
$j=-\rho_0\dot{\varphi}/Q $.

Because $4k_F$ modulations of SDWs or CDWs are also
possible~\cite{Tomio00}, we introduce the factor $p$ in the
argument of the modulating cosine function, i.e., for CDWs and
SDWs $p$ is usually $1$, but can also be $2$ or greater.

\subsection{Hamiltonian}

In the following we use a minimal model for the low energy, long
wave length excitations of the condensed charge density wave.
Since fluctuations in the amplitude $|\Delta|$ are suppressed,
because these are massive, we take into account only fluctuations
in the phase $\varphi$. Clearly, such an approach breaks down
sufficiently close to the mean-field transition temperature
$T_c^{\textit{MF}}$. Neglecting fluctuations in $|\Delta|$, the
Hamiltonian of the CDW is given by
\begin{equation}
 \hat {\cal H}  =\hat {\cal H}_0+\hat {\cal H}_u+\hat {\cal H}_w\label{eq.qham}
\end{equation}
with
\addtocounter{equation}{-1}
\begin{subequations}
\begin{eqnarray}
\hat {\cal
H}_0&\equiv&\int\limits_{0}^{L}dx\,\frac{c}{2}\left[\left(\frac{v}{c}
\right)^2\hat P^2+(\partial_x\hat\varphi)^2\right]\,,\label{eq.H0}\\
\hat {\cal H}_u&\equiv&\int\limits_{0}^{L}dx\, U(x)\rho(x)\;,\; U(x)=\sum_{i=1}^{N_{\textit{imp}}} U_{i}\delta(x-x_i)\label{eq.Hu}\\
\hat {\cal H}_w&\equiv&-\int\limits_{0}^{L}dx\,W\cos
\big(q\hat\varphi(x)\big)\,.\label{eq.Hw}
\end{eqnarray}
\end{subequations}
$\hat {\cal H}_0$ describes the phason excitations of the CDW,
where $c=\frac{\hbar v_F}{2\pi} f(T)$ denotes the elastic
constant. $v=v_F/\sqrt{1+(2|\Delta|/\hbar\omega_{pQ})^2/(gf)}$ is
the effective velocity of the phason excitations with
$\omega_{pQ}$ the phonon frequency. For CDWs
$(2|\Delta|/\hbar\omega_{pQ})^2/(gf) \gg 1$ is typically fulfilled
and hence quantum fluctuations are weak.

$\hat P$ is the momentum operator, corresponding to the phase
$\hat\varphi$, with the standard commutation relation $[\hat
P(x),\hat\varphi(x^{\prime})]=\frac{\hbar}{i}\delta(x-x^{\prime})$

$\hat {\cal H}_u$ results from the effects of impurities with
random potential strength $U_i$ and positions $x_i$. The potential
strength  is characterized by $\dav{U_i}=0$ and
$\dav{U_iU_j}\equiv U_{\textit{imp}}^2\delta_{i,j}$,  and includes
a forward and a backward scattering term proportional to $\rho_0$
and $\rho_1$, respectively. The disorder average of the impurity
potential $U(x)$ follows then to be given by $\dav{U(x)}=0$ and
\begin{equation}\label{eq.UUMW}
\dav{U(x)U(y)}=
{\frac{U_{\textit{imp}}^2}{l_{\textit{imp}}}}\delta(x-y)\,.
\end{equation}

We will further assume, that the mean impurity distance
$l_{\textit{imp}}=L/N_{\textit{imp}}$ is large compared with the
wave length of the CDW and, in most parts of the paper,  that the
disorder is weak, i.e.,
\begin{equation}\label{eq.disorderstrength}
  1\ll l_{\textit{imp}}Q \ll c Q/(U_{\textit{imp}}\rho_1).
\end{equation}
In this case the Fukuyama--Lee length
\begin{equation}\label{eq.FL}
L_{\textit{FL}}=\left(\frac{c\sqrt
{l_{\textit{imp}}}}{U_{\textit{imp}}\rho_1p^2}\right)^{2/3}
\end{equation}
is large compared to the impurity distance $l_{\textit{imp}}$.

The third term in (\ref{eq.qham}), ${\cal H}_w$, includes the
influence of a harmonic lattice potential. This term will be
discussed section \ref{sec.lp} in greater detail.

Our model (\ref{eq.qham}) includes the four dimensionless
parameters
\begin{subequations}
\begin{eqnarray}
t&=&T/\pi\Lambda c\,,\label{eq.t}\\
K&=&\hbar v /\pi c\,,\label{eq.K}\\
u^2&=&\frac{(U_{\textit{imp}}\rho_1)^2}{\Lambda^3\pi c^2 l_{\textit{imp}}}\,,\label{eq.u}\\
w&=&W/\pi c\Lambda^2\,,\label{eq.w}
\end{eqnarray}
\end{subequations}
which measure the strength of the thermal ($t$), quantum ($K$) and
disorder fluctuations ($u$) and the periodic potential ($w$),
respectively. $\Lambda=\pi/a$ is a momentum cut-off. The classical
region of the model is given by $K\ll t$ which can be rewritten as
the condition, that the thermal de Broglie wave length
\begin{equation}\label{eq.lambdaT}
  \lambda_T=\hbar\beta v=K/(t\Lambda)
\end{equation}
of the phason excitations is small compared to $a$.

At $T=0$,  $K$-values of the order $10^{-2}$ to $10^{-1}$ and $1$,
have been discussed for CDWs and SDWs,
respectively~\cite{Gruener,Maki}. It has to be noted however, that
the the expressions relating $c$ and $v$ to the microscopic
(mean-field-like) theory lead to the conclusion that $K$ and $t$
diverge by approaching $T_c^{\textit{MF}}$, whereas the ratio
$K/t$ remains finite.

\subsection{Coulomb Interaction}

We could also add a {\it Coulomb interaction} term to our model
(\ref{eq.qham}) which can be written as
\begin{equation}
\hat {\cal H}_c=\frac{1}{2}\int\, dx\int\,dx^{\prime}
\hat\rho(x)V_c(x-x^{\prime})\hat\rho(x^{\prime})\,,\label{eq.HC}
\end{equation}
where $V_c$ is the Coulomb potential. In all dimensions the
unscreened potential has the form $e^2/r$. If we assume, that the
quasi one-dimensional system has the finite width $\zeta$, $V_c$
can be written as~\cite{Maurey95,Schulz93}
\begin{eqnarray}
 V_c^0(x)&=&\frac{e^2}{\sqrt{x^2+\zeta^2}}=\frac{1}{L}\sum_{k}e^{\imath
 kx}V_c^0(k)\,\,\text{with}\\
 &&V_c^0(k)=2e^2K_0(|\zeta k|)\,,\label{eq.Vc0k}
\end{eqnarray}
where $K_0$ is a Bessel function with $K_0(x)\approx -\ln(x)$ for
$x\ll 1$.

In general the Coulomb potential is screened and can be written
as~\cite{Altshuler85}
\begin{equation}
 V_c(k,\omega)=\frac{V_c^0(k)}{1+V_c^0(k)\Pi(k,\omega)\,,}
\end{equation}
with the momentum and frequency dependent polarization operator
$\Pi(k,\omega)=\tav{\rho(0,0)\rho(k,\omega)}$.

If we only consider the static case $\omega=0$ we can distinguish
two limiting cases: First, if the typical range
$\lambda_{\textit{eff}}$ of the screened Coulomb potential $V_c$
is much smaller than the mean electron distance, the potential can
be assumed to be a delta distribution and $H_c$ can be
approximated by
\begin{equation}
\hat {\cal H}_c\approx\frac{\hbar \chi}{2}\int\, dx
\left(\frac{1}{\pi}\partial_x\varphi(x)\right)^2+\ldots\,,\label{eq.HC2}
\end{equation}
with $\chi=\frac{1}{\hbar}\int\, dx V_c(x)$. The $\cos$-terms
($\ldots$) from the density can be neglected due to strong
fluctuations. Therefore the Coulomb interaction gives only an
additional contribution to the elastic constant of the initial
model: $c=\frac{\hbar v_F}{2\pi}f+\frac{\hbar \chi}{\pi^2}$. For
$\chi>0$ the Coulomb interaction is repulsive, which leads to an
increase of $c$ and therefore a decrease of the dimensionless
parameter $K$, i.e., the quantum fluctuations will be reduced by
the Coulomb interaction. In the  case $\chi<0$ (attraction), $K$
will be increased. Keeping this consideration in mind, we will not
further include $\hat{\cal H}_c$ in the model explicitly.

In the other case -- with weak screening -- $V_c(k)\approx
V_c^0(k)$ shows the dispersion given in (\ref{eq.Vc0k}) and in
general, the details of the $k$-dependence depend not only on the
transverse extension $\zeta$ of the quasi one-dimensional system
under consideration but also on the screening
length~\cite{FisherGrinstein,Maurey95}.

However, the logarithmic $k$-dependence will only weakly affect
our RG-analysis, but may suppress phase transitions, as discussed
later in section \ref{ssec.zero}.

Coulomb interaction is also important if one considers
multi-channel systems\cite{Lee78} or the effect of the
non--condensated normal electrons.

\section{Renormalization group treatment of disorder}\label{sec.disorder}

\subsection{Flow equations}

In order to determine the phase diagram we adopt a standard
Wilson-type renormalization group calculation, which starts from a
path integral formulation of the partition function corresponding
to the Hamiltonian (\ref{eq.qham}). We begin with the
renormalization of the disorder term and put $w=0$ in the
following. The system is transformed into a translational
invariant problem using the replica method, in which the disorder
averaged free energy is calculated, using
\begin{equation}\label{eq.F}
\dav{{\cal F}}=-T\dav{\ln\Tr e^{-{\cal
S}/\hbar}}\equiv-T\lim_{n\rightarrow 0}\frac{1}{n}\left(\Tr
e^{-{\cal S}^{(n)}/\hbar}-1\right)\,,
\end{equation}
which defines the replicated action ${\cal S}^{(n)}$. ${\cal
S}^{(n)}$ is given by
\begin{equation}\label{eq.Sn}
{\cal S}^{(n)}=\sum_{\alpha,\nu}\int_{\tau}\,\left\{{\cal
L}_{0,\alpha}\delta_{\alpha\nu}+\frac{1}{2\hbar}\int_{\tau^{\prime}}\,\dav{{\cal
H}_{u}[\varphi_{\alpha}(\tau)]{\cal
H}_{u}[\varphi_{\nu}(\tau^{\prime})]}\right\}\,,
\end{equation}
where ${\cal L}_0$ is the Lagrangian corresponding to $\hat {\cal
H}_0$, $\int_{\tau}\equiv\int_0^{\hbar\beta}d\tau$ and
$\alpha,\nu$ are replica indices. Using (\ref{eq.UUMW}) and
consequently neglecting higher harmonics ($2pQ$-modes) one
finds

\begin{widetext}

\begin{equation}
\dav{{\cal H}_{u}[\varphi_{\alpha}(\tau)]{\cal
H}_{u}[\varphi_{\nu}(\tau^{\prime})]}=\frac{
U_{\textit{imp}}^2\rho_1^2}{2l_{\textit{imp}}}\int\limits_0^Ldx\,\left\{\cos
p\Big(\varphi_{\alpha}(x,\tau)-\varphi_{\nu}(x,\tau^{\prime})\Big)
+\frac{2\rho_0^2}{Q^2\rho_1^2}\partial_x\varphi_{\alpha}(x,\tau)\partial_x\varphi_{\nu}(x,\tau^{\prime})
\right\}\,.
\end{equation}
Together with (\ref{eq.Sn}) one obtains the following form

\begin{eqnarray}
\frac{{\cal S}^{(n)}}{\hbar}  =  \frac{1}{2\pi K}\sum_{\alpha,\nu}
\int\limits_0^{L\Lambda}dx\int\limits_0^{K/t}d\tau\Bigg\{&&
\Big[(\partial_x\varphi_{\alpha})^2+(\partial_{\tau}\varphi_{\alpha})^2\Big]
\delta_{\alpha\nu} -
\frac{1}{2K}\int\limits_0^{K/t}d\tau^{\prime}\Big[u^2\cos{p}
\Big(\varphi_{\alpha}(x,\tau)-\varphi_{\nu}(x,\tau^{\prime})\Big)\nonumber\\
&&+\sigma\partial_x\varphi_{\alpha}(x,\tau)\partial_x\varphi_{\nu}(x,\tau^{\prime})\Big]\Bigg\}\,,
\label{eq.S^n/hbar}
\end{eqnarray}
with $\sigma=2u^2(\rho_0\Lambda/\rho_1Q)^2$.

\end{widetext}
Note, that we introduced  dimensionless spatial and imaginary time
variables,
\begin{eqnarray}
\Lambda x \rightarrow x\,,\nn\\
\Lambda v\tau \rightarrow \tau\,,\nn
\end{eqnarray}
which will be used throughout the paper - beginning here.
Furthermore all lengths (e.g. correlation lengths, $\lambda_T$,
$L_{\textit{FL}}$, $l_{\textit{imp}}$ and $L$), wave vectors (e.g.
$k$, $k_F$ and $Q$) and Matsubara frequencies are dimensionless
accordingly, from now on. Additionally we rescale the elastic
constant
\begin{equation*}
\Lambda c \to c\,,
\end{equation*}
for convenience to avoid the appearance of $\Lambda$.

Integrating over the high momentum modes of $\varphi(x,\tau)$ in a
momentum shell of infinitesimal width $1/b\le|q|\le 1$ but
arbitrary frequencies and rescaling $x\rightarrow x'=x/b$, $\tau
\rightarrow \tau'=\tau/b$, we obtain the following renormalization
group flow equations (up to one loop):
\begin{subequations}
\begin{eqnarray}
\frac{dt}{dl}  &=&t\,,\label{eq.dt/dl}\\
\frac{dK}{dl} & = & -
\frac{1}{2}p^4u^2KB_0\left(p^2K,\frac{K}{2t}\right)
\coth \frac{K}{2t}\label{eq.dK/dl},\\
\frac{du^2}{dl} & = & \Big[3  -
\frac{p^2K}{2}\coth{\frac{K}{2t}}\Big]u^2\,,\label{eq.du/dl}\\
\frac{d\sigma}{dl}&=&\sigma\,,\label{eq.ds/dl}
\end{eqnarray}
\end{subequations}
where $l=\ln{b}$. For details on the RG calculation we relegate to
Appendix \ref{app.RG} where we have written the RG-flow also for
dimensions $0<|d-1|\ll 1$.

For legibility we have introduced the following functions:
\begin{eqnarray}
B_i(\nu,y) &=& \int\limits_0^y d\tau\,\int\limits_0^\infty dx\,
  \frac{g_i(\tau,x)}{\Upsilon(\tau,x)}\frac{\cosh{(y-\tau)}}{\cosh{y}},\label{eq.B}\\
\Upsilon(\tau,x )&=&\left[1+\left(\frac{y}{\pi}\right)^2
  \left(\cosh{\frac{\pi
  x}{y}}-\cos{\frac{\pi\tau}{y}}\right)\right]^{\nu/4}\,,\label{eq.Y}
\end{eqnarray}
with
\begin{eqnarray}
 g_0(\tau,x)&=&\delta (x)\tau^2\,,\nn
\end{eqnarray}
Note, that $B_0(p^2K,\frac{K}{2t})\rightarrow 0$ for $K
\rightarrow 0$ (see Fig. \ref{figB0} in Appendix \ref{app.RG}).

There is no renormalization of $t$ (i.e., of the elastic constant
$c$) because of a statistical tilt symmetry~\cite{Schultz88}. Note
that (\ref{eq.S^n/hbar}) is written in rescaled dimensionless
parameters and the different renormalization of the kinetic and
elastic term is reflected in the different renormalization of $v$
and $c$, i.e., $K$ and $t$, respectively.

From the flow equation for $u^2$ (\ref{eq.du/dl}) one sees
directly that, depending on the sign of the prefactor, the
behavior changes from increase for small $t$ and $K$ to decrease
for high $K$ or $t$.


There is no first order RG correction to $\sigma$ and the change
of $\sigma$ with length scale is simply given by rescaling, see
(\ref{eq.ds/dl}). The two-loop contribution to $\sigma$ is much
more involved than the one-loop contributions for the other flow
equations and gives no qualitatively different result for the flow
of $\sigma$. As seen from (\ref{eq.ds/dl}), the forward scattering
amplitude always increases as $\sigma_0 e^l$ on larger length
scales and is therefore not well controlled in the RG sense. But,
since the flow of $\sigma$ does not feed back into the other flow
equations it has only minor relevance for our considerations. And
indeed, we can get rid of the forward scattering term $f/\pi
U(x)\frac{\partial \varphi}{\partial x}$  by introducing the field
$\hat\varphi_b(x)$ by
\begin{equation}\label{eq.varphide}
\hat\varphi(x)=\hat\varphi_b(x)-
\varphi_{f}(x)\;,\quad\varphi_{f}(x)=\int^{x}_0 dy c(y)\,,
\end{equation}
where $c(x)\equiv\frac{U(x)f}{\pi c\Lambda}$, with $\dav{c(x)}=0$
and $\dav{c(x)c(x')} =\frac{\pi}{2}\sigma \delta (x-x')$. This can
be easily verified by inserting this decomposition of
$\hat\varphi(x)$ into the initial Hamiltonian (\ref{eq.qham})
written in dimensionless units, and using (\ref{eq.UUMW}) and the
definition of $\sigma$ for deriving the averages of $c(x)$. Note,
that $x$ is dimensionless.


\begin{table}[htb]
\caption{\label{tab.nguide} Notation guide.}
\begin{ruledtabular}
\begin{tabular}{ccc}
symbol here& G. \& S.\footnote{charge operators}, Ref. [\onlinecite{GiaSchulz87}]& Haldane, Ref. [\onlinecite{Haldane}]\\
\hline
$\varphi$, $\hat j$ & $\sqrt{2}\phi$, $\sqrt{2}/\pi\partial_{\tau}\phi$ & $\theta-\pi\rho_0 x$, $\pi^{-1}\dot\theta$\\
$\hat P$ & $\hbar\Pi/\sqrt{2}$ & $-\frac{\hbar}{\pi}\nabla\varphi$\\
$K$ & $2K_{\rho}$ & $\sqrt{v_j/v_N}$\\
$v$ & $u_{\rho}$ & $\sqrt{v_j v_N}$\\
$c$ & $\frac{\hbar u_{\rho}}{2\pi K_{\rho}}$ & $\hbar v_N/\pi$\\
$p$ & $1$ & $2$\\
\end{tabular}
\end{ruledtabular}
\end{table}

\subsection{Zero temperature - a review}\label{ssec.zero}

The special case $t=0$ was previously considered, e.g., in
[\onlinecite{GiaSchulz87}] (for a better comparison see our {\it
notation guide} given in table \ref{tab.nguide}).

The flow equations for $K$ and $u$ at zero temperature read:
\begin{subequations}
\begin{eqnarray}
\frac{dK}{dl} & = & -  \frac{1}{2}p^4u^2KB_0(p^2K,\infty)\,,\label{eq.flowK0}\\
\frac{du^2}{dl} & = & \Big[3  -
\frac{p^2K}{2}\Big]u^2\,,\label{eq.flowu0}
\end{eqnarray}
\end{subequations}
with
\begin{equation}
B_0(\nu,\infty)=\int\limits_0^\infty d\tau\, \tau^2
e^{-\tau}\left[1+\tau^2/2\right]^{-\nu/4}\,.\label{eq.B00}
\end{equation}

The corresponding flow equation for $K$ obtained in
[\onlinecite{GiaSchulz87}] deviates slightly from
(\ref{eq.flowK0}), which can be traced back to the different RG
procedures. In  [\onlinecite{GiaSchulz87}] the authors performed
the RG at strictly zero temperature and used a symmetric, circular
shape of the ''momentum--shell'', i.e., treated the model as
effectively isotropic in the 1+1-dimensional space-time.

This procedure may be a good approximation at zero temperature,
but if one considerers finite temperatures this does not hold
anymore, since the extension in  $\tau$-direction is now finite.
As a result, there is a region $\pi/L<|k|<\pi/\lambda_T$ where
fluctuations are mainly one-dimensional and purely  thermal. This
region was disregarded in previous treatments. As we will see,
fluctuations from this  region have an important influence on the
overall phase diagram.

The critical behavior is, however, the same: there is a
Kosterlitz-Thouless (KT) transition at the phase boundary $K_u$
between a disorder dominated, pinned and a free, unpinned phase
which terminates in the fixed point $K_u^*=6/p^2$. One can derive
an implicit equation for $K_u$ by combining (\ref{eq.flowK0}) and
(\ref{eq.flowu0}) to a differential equation
\begin{equation}\label{eq.du/dK}
\frac{du^2}{dK}=\frac{1}{p^2\eta K}(K-K_u^*)\,,
\end{equation}
which has the solution
\begin{equation}\label{eq.du/dK.sol}
u^2(K)-u_0^2=\frac{K_u^*}{p^2\eta}\left(\frac{K-K_0}{K_u^*}-\ln\frac{K}{K_0}\right)\,,
\end{equation}
where $u_0$ and $K_0$ denote the bare values of the disorder and
quantum fluctuation, respectively, and $\eta\equiv
B_0(p^2K_u^*,\infty)$. Then, $K_u$ is implicitly given by
\begin{equation}\label{eq.Ku}
u^2(K_u)=\frac{K_u^*}{p^2\eta}\left(\frac{K_u-K_u^*}{K_u^*}-\ln{\frac{K_u}{K_u^*}}\right)\,,
\end{equation}
where the initial condition $u^2(K_0=K_u^*)=u_0^2=0$ is used. The
KT-flow equations at $K_u^*$ can be recovered by defining
\begin{eqnarray}
  2\gamma&\equiv& \frac{p^2K}{2}-3\,,\nn\\
  2\chi^2&\equiv& \frac{3}{2}p^4\eta u^2\nn
\end{eqnarray}
with $|\gamma|\ll 1$. This yields
\begin{subequations}
\begin{eqnarray}
 \frac{d\gamma}{dl}&=&-\chi^2\; ,\label{eq.KTgamma}\\
 \frac{d\chi^2}{dl}&=&-2\gamma\chi^2\,,\label{eq.KTchi}
\end{eqnarray}
\end{subequations}
which are exactly the flow equations obtained by Kosterlitz and
Thouless~\cite{Kosterlitz}.

\begin{figure}[htb]
\includegraphics[width=0.6\linewidth]{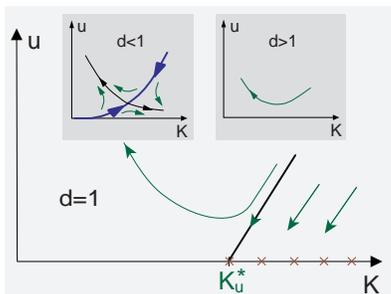}
\caption{Schematic zero temperature phase diagram in  $d=1$ and
close to $d=1$ dimensions (see text). $u$ and $K$ denote the
strength of the disorder and quantum fluctuations, respectively.
}\label{fig.closed1}
\end{figure}

Under the assumption, that a small deviation from the dimension
$d=1$ changes only the naive scaling dimensions of the fields, our
results can be extended also to $d=1+\epsilon$ dimensions (For
details see Appendix B). The zero temperature phase diagram is
modified and illustrated in Fig. \ref{fig.closed1}. For
$\epsilon<0$ the fixed point at ($K=K_u^{\ast}$, $u=0$) is shifted
to positive $u$-values (see left inset of Fig. \ref{fig.closed1}),
whereas for $\epsilon>0$, $K$ and $u$ always flow to the strong
pinning fixed point (at $K=0$ and $u\rightarrow\infty$; right
inset), i.e., quantum fluctuations are too weak to renormalize the
random potential to zero.  The zero temperature transition
disappears therefore for $d>1$, since the fixed point lies in the
unphysical $u<0$ region of the $K$-$u$ parameter space. This can
easily be verified by using the rescaling of $K$ given in eq.
(\ref{eq.rescaleK}) of Appendix \ref{app.RG} which results in the
flow equation (\ref{eq.flowK}). In general this discussion applies
to the localization transition as well as to the Mott transition
(see discussion of the the lattice potential). Note, that the flow
for $d \ne 1$ is qualitatively different from that discussed in
Ref. [\onlinecite{Herbut99}].

If one includes the effect of \emph{Coulomb interaction} in $d=1$
dimension, phase fluctuations of the free phase field increase
only as (T=0)
\begin{equation}\label{eq.CFcoulomb}
  \tav{(\varphi(x,0)-\varphi(0,0))^2}\sim K\ln^{1/2}|x|.
\end{equation}
As a result, phase fluctuations are too weak to suppress the
disorder even for large values of $K$ and the system is always in
the pinned phase. The phase diagram is therefore similar to that
in $d>1$ dimensions.

In the pinned phase the parameters $K$ and $u$ flow into the
classical, strong disorder region: $K\rightarrow 0, u\rightarrow
\infty$.

Integration of the flow equations gives for small initial disorder
and $K\ll K_u^{\ast}$ an effective correlation length or
localization length
\begin{equation}\label{eq.xiu}
\xi_{u} \approx L_{\textit{FL}}^{(1-K/K_u^{\ast})^{-1}}\,,
\end{equation}
at which $u$ becomes of the order unity. This can be extracted
from (\ref{eq.flowu0}) neglecting the flow of $K$.

A better approximation of $\xi_u$ which takes also the flow of $K$
into account can be obtained by replacing $u^2$ in the flow
equation for $K$ (\ref{eq.flowK0}) by the expression given in
(\ref{eq.du/dK.sol}). We still use the approximation, that $K$
deviates not much from the bare value $K_0$ which is the case, as
long as $u_0^2 l\ll 1$. Then, the solution for $K(l)$ is given by
\begin{equation}
K(l)\approx K_0\left(1-\frac{p^4}{2}u_0^2\eta l\right)\,,
\end{equation}
which yields a solution of (\ref{eq.flowu0}):
\begin{equation}
\ln\frac{u^2(l)}{u_0^2}\approx\left(3-\frac{p^2}{2}K_0\right)l+\frac{p^6}{8}\eta
K_0u_0^2 l^2\,.
\end{equation}
With $u^2(\ln(\xi_u))\approx 1$, the correlation length $\xi_u$ is
defined by
\begin{equation}
0=\ln u_0^2+\underbrace{\left(3-\frac{p^2}{2}K_0\right)}_{\equiv
a}\ln(\xi_u)+\underbrace{\frac{p^6}{8}\eta K_0u_0^2}_{\equiv
b}(\ln(\xi_u))^2
\end{equation}
which yields
\begin{eqnarray}
\ln(\xi_u)&=&\frac{\sqrt{a^2-4b\ln u_0^2}-a}{2b}\nn\\
&\approx& -\frac{\ln u_0^2}{3-\frac{p^2K_0}{2}}-\frac{p^6}{8}\eta
K_0u_0^2\frac{(\ln u_0^2)^2}{\left(3-\frac{p^2K_0}{2}\right)^3}\,,
\end{eqnarray}
where the first term of the r.h.s. gives the result
(\ref{eq.xiu}).

Close to the transition line $\xi_u$ shows KT behavior. For $K \ge
K_u$, $\xi_u$ diverges and $C(x,\tau)\sim K(l=\ln|z|)\ln|z|$ where
$|z|=\sqrt{x^2+\tau^2}$ (cf. section \ref{sec.cf}).

\subsection{Strong pinning limit: Exact ground state}\label{ssec.strong}

For large values of $u$ our flow equations break down.
Qualitatively the flow is towards large $u$ and small $K$.  We
can, however, find the \emph{asymptotic} behavior in this phase by
solving the initial model in the {\it strong pinning limit
exactly}. To find this solution we will assume strong pinning
centers and weak thermal fluctuations:

\begin{equation}\label{eq.strong.pinning.condition}
U_{\textit{imp}}\to\infty\,\,\,\,\text{and}\,\,\,\,
c/l_{\textit{imp}}\gg T.
\end{equation}

To treat this case we go back to the initial Hamiltonian
(\ref{eq.qham}) (with $W\equiv 0$ and the kinetic term also
vanishes because of $K\to 0$). For strong disorder it is
convenient to integrate out the phase field $\varphi(x)$ at all
points which are not affected by the impurities. Then the
effective Hamiltonian takes the form~\cite{Feigel80}
   \begin{equation}
   {\cal H}_{\textit{eff}}=\sum_{i=1}^{N}
     \left\{ \frac{c}{2}\frac{(\varphi_{i+1}-\varphi_i)^2}{x_{i+1}-x_i}
    + U_{{i}} \rho(x_i)\right\}\;,\; \varphi_i\equiv\varphi(x_i)\,.
   \label{eq.strongham}
   \end{equation}
Under condition (\ref{eq.strong.pinning.condition}), $\varphi_i$
only takes values obeying
\begin{equation}\label{eq.strong.pinning.condition2}
p(\varphi_i+Qx_i)=2\pi n_i+\pi\,\,\, \text{with} \,\,\, n_i \in
\mathbb{Z} \,\,\,\text{integer}
\end{equation}
which minimizes the backward scattering term. Defining $h_i$ and
$\epsilon_i$ by
\begin{equation}\label{eq.hi}
  n_{i+1}-n_i\equiv h_i+\gb{\frac{p Q
  l_{\textit{imp}}}{2\pi}},\,\,\,\,
x_{i+1}-x_i\equiv l_{\textit{imp}}+\epsilon_i
\end{equation}
 with $0\le x_1\le x_2\le\ldots\le x_{N+1}\le L$, the effective Hamiltonian can be
rewritten as
\begin{equation}
{\cal H}_{\textit{eff}}=\frac{c}{2p^2}\sum_{i}
\frac{(2\pi)^2\left(h_i-\frac{pQ\epsilon_i}{2\pi}-
\gamma\right)^2}{l_{\textit{imp}}+\epsilon_i}\,.
\end{equation}
Here $\gb{x}$ denotes the closest integer to $x$ ({\it Gaussian
brackets}):
\begin{equation}
\gb{x}=m\,\,\,\,\text{for}\,\,\,\,x\in \left]m-\frac{1}{2} ,
m+\frac{1}{2}\right]\,\,, m\in\mathbb{Z}
\end{equation}
and
\begin{equation}\label{eq.spgamma}
\gamma\equiv\frac{pQl_{\textit{imp}}}{2\pi}-\gb{\frac{pQl_{\textit{imp}}}{2\pi}},
\end{equation}
such that $|\gamma|\le\frac{1}{2}$.

Because thermal fluctuations are small compared to the elastic
energy, see (\ref{eq.strong.pinning.condition}), $(h_i-\frac{p
Q\epsilon_i}{2\pi}-\gamma)$ takes on its minimal value, which is
given by
\begin{equation}\label{eq.himin}
  h_i^0=\gb{\frac{pQ\epsilon_i}{2\pi}+\gamma}.
\end{equation}
which defines the exact ground state of the classical model: If we
use (\ref{eq.hi}) one finds for the optimal value of the $n_i$'s
\begin{equation}
n_{i+1}^0=n_i^0+\gb{\frac{pQ}{2\pi}(\epsilon_i+l_{\textit{imp}})}\nonumber\\
\end{equation}
which leads, using (\ref{eq.strong.pinning.condition2}), to the
exact classical ground state
\begin{equation}
 \varphi_i^0=\frac{1}{p}\left(2\pi\left\{n_1^0+\sum_{j<i}\gb{\frac{pQ}{2\pi}(\epsilon_j+l_{\textit{imp}})}\right\}+\pi\right)-Qx_i\,,
\end{equation}
where $n_1^0$ has an arbitrary integer value.


\subsection{Finite temperature and crossover diagram}\label{ssec.crossover}


\begin{figure}[htb]
\includegraphics[width=0.6\linewidth]{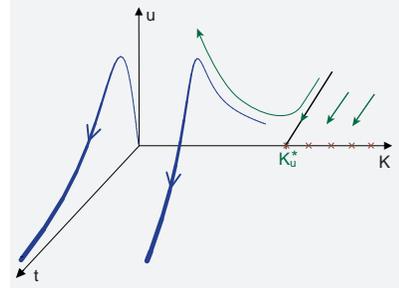}
\caption{Typical flow diagram for $w=0$ in the three dimensional
parameter space of $K$, $u$ and $t$, proportional to the strength
of quantum, disorder and thermal fluctuations, respectively.
}\label{fig.flowu}
\end{figure}

At {\it finite temperatures} thermal fluctuations wipe out the
random potential which lead to the pinning of the CDW at $t=0$ and
$K<K_u$. Thus there is no phase transition any more, in agreement
with the Landau theorem. The system is always in its delocalized
phase even if the disorder may still play a significant role on
intermediate length scales.

In the special case $K\to 0$ the flow equation (\ref{eq.du/dl})
reduces to $\frac{du^2}{dl} = \Big[3 -p^2 t\Big]u^2$ with solution
\begin{equation}
u^2(l)=u_0^2e^{3l-p^2t_0(e^l-1)}\,.\label{eq.ul_K0}
\end{equation}
If we write $t=t_0 e^l$ we may express $l$ by $t$ and hence we may
write $u^2$ as $t$-dependent function:
\begin{equation}
u^2(t)=u^2_0(t/t_0)^3e^{-p^2(t-t_0)}\,,
\end{equation}
which is plotted in Fig. \ref{fig.flowu} in the $t$-$u$ plane.

One sees that the flow of the disorder has a maximum at $t=3/p^2$
or $l=\ln(3/(p^2t_0))$, if $t_0<3/p^2$. For finite $K$, the RG
flow of $u$ in the region $K < K_u$ first increases and then
decreases. The region of increase in the $K$-$t$ plane is
implicitly defined by ${\cal M}_u\equiv
\left\{(K,t)|K_u^{\ast}\geq K\coth\frac{K}{2t}\geq 0\right\}$,
i.e., the positions of the maxima of $u^2[K,t]$ are located on the
boundary of ${\cal M}_u$ defined by $K_u^{\ast}=
K\coth\frac{K}{2t}$.

\begin{figure}[htb]
\includegraphics[width=0.95\linewidth]{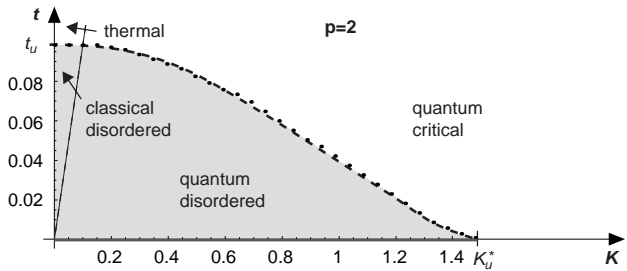}
\caption{The low temperature crossover diagram of a
one-dimensional CDW. $t$ and $K$ are proportional to the
temperature and the strength of quantum fluctuations,
respectively. The amount of disorder corresponds to a reduced
temperature $t_{\textit{u}}\approx 0.1$. In the classical and
quantum disordered region, respectively, essentially the $t=0$
behavior is seen. The straight dashed line separating them
corresponds to $\lambda_T\approx 1$, i.e., $K\approx t$. In the
quantum critical region the correlation length is given by
$\lambda_T$. Pinning (localization) occurs only for $t=0,
K<K_u^*$.}\label{fig.crossover}
\end{figure}

The correlation length $\xi$ can be found approximately by
integrating the flow equations until the maximum of $u(l)$ and
$t(l)/(1+K(l))$ is of the order one (see discussion in section
\ref{sec.cf}). This can be done in full generality only
numerically (see Fig. \ref{fig.crossover}).

It is however possible to discuss several special cases
analytically.  The zero temperature correlation length can still
be observed as long as this is smaller than the thermal de Brogle
wave length $\lambda_T$ which can be rewritten for $K$ not too
close to $K_u^*$ as $t\lesssim t_K\approx
Kt_{\textit{u}}^{(1-K/K_u)^{-1}}$ with $t_{u}\approx
L_{\textit{FL}}^{-1}$, where we defined $t_K$ via $\xi_u\equiv
\frac{K}{t_K}$, analogously to the definition of $\lambda_T$, and
used (\ref{eq.xiu}). We call this domain the {\it quantum
disordered region}.

For $K\ge K_u$ the correlation length $\xi$ is given by
$\lambda_T$ which is larger than given by purely thermal
fluctuations. For scales smaller than $\lambda_T$, the phase
correlation function still increases as $\sim \ln |z|$ with a
continuously varying coefficient $K_{\textit{eff}}(u_0)$, as will
be discussed in detail in the next section. In this sense one
observes {\it quantum critical behavior} in that region, despite
of the fact, that the correlation length is now finite for all
values of $K$ [\onlinecite{Chakra}].

In the {\it classical disordered region} $t_K<t<t_{u}$ the
correlation length is roughly given by $L_{\textit{FL}}$ as
follows from previous studies~\cite{Feigel80,ViFer84} or by
solving  $u^2(\ln(\xi))\simeq 1$ using (\ref{eq.ul_K0}) for small
$t_0$ yielding $\xi\approx u_0^{-2/3}e^{-p^2t_0}\approx
u_0^{-2/3}= L_{\textit{FL}}(\pi p^4)^{1/3}$. Note, that
$t_K\approx K$ for small $K$.

In the remaining region $t \gtrsim t_{u}$, the {\it thermal
region}, we apply the mapping onto the Burgers equation (see
section \ref{sec.cf}). In this case the RG-procedure applied to
this equation becomes trivial since there is only a contribution
from a single momentum shell and one finds for the correlation
length $\xi^{-1}\approx \frac{\pi}{2}f(T)t[1+1/2[t_{u} /(\pi p^2
t)]^3]\Lambda$.

The phase diagram depicted in Fig. \ref{fig.crossover} is the
result of the numerical integration of our flow equations and
shows indeed the various crossovers discussed before.

In the {\it high temperature} region ($t\gg K$) the flow equations
can be solved explicitly. For $u^2(l)$ we get the same result as
given in (\ref{eq.ul_K0}) and the flow equation for $K$ reduces to
\begin{equation}
 \frac{dK}{dl}=-\frac{p^4}{2}u^2\frac{K^4}{(2t)^3}\,,
\end{equation}
where we used $B_0(p^2K,\frac{K}{2t}\to 0)=(K/2t)^4$. The solution
of this equation is given by
\begin{equation}
K(l)=\left[K_0^{-3}+\frac{3p^4u_0^2}{16t_0^3}e^{p^2t_0}\Ei(p^2t_0,p^2t_0e^l)\right]^{-1/3}\,,
\end{equation}
with the incomplete exponential integral function $\Ei(a,b)$
defined by
\begin{equation*}
\Ei(a,b)\equiv \int_{a}^{b}dt\,e^{-t}/t\,.
\end{equation*}
One observes that $K(l)$ saturates very quickly to the value
$K(\infty)<K_0$.


\section{Correlation functions}\label{sec.cf}


In this section we discuss the density-density and the phase
correlation functions in more detail and summarize all correlation
lengths in the various regimes -- partly already used in the last
two sections.

The (full) density-density correlation function is defined as
\begin{equation}\label{eq.densityCF}
 S(x,\tau)\equiv\tav{\rho(x,\tau)\rho(0,0)}\,,
\end{equation}

where $\rho(x,\tau)$ is given in (\ref{eq.cdwdensity}). In the
following we restrict our considerations to the {\it (charge)
density wave order} part of $S$, which is the term proportional to
$\rho_1^2$, i.e.
\begin{equation}\label{eq.densityCFcdw}
S_1(x,\tau)=\rho_1^2\tav{\cos p(\varphi(x,\tau)+Qx)\cos p
\varphi(0,0)}\,,
\end{equation}
which defines the type of order of the density wave: If it decays
algebraically we have quasi long--range order (QLRO), an
exponential decay  over a correlation length $\xi$ corresponds to
short--range order (SRO). The omitted  parts of $S$ decay faster
than $S_1$ [\onlinecite{GiaSchu89}].

$S_1$ can be rewritten as
\begin{eqnarray}
S_1(x,\tau)=\frac{\rho_1^2}{4}\Big(&&e^{\imath p Q
x}\tav{e^{\imath p(\varphi(x,\tau)-\varphi(0,0))}}+\nn\\
&&e^{-\imath p Q x}\tav{e^{-\imath
p(\varphi(x,\tau)-\varphi(0,0))}}\Big)\,,
\end{eqnarray}
and under the assumption that we can use a gaussian approximation
for the averages, we obtain
\begin{equation}\label{eq.densityCFcdw.exp}
S_1(x,\tau)\simeq
\rho_1^2\cos(pQx)e^{-\frac{p^2}{2}\tav{(\varphi(x,\tau)-\varphi(0,0))^2}}\,.
\end{equation}

From now on we focus on the \emph{phase correlation function}
\begin{equation}\label{eq.phaseCF}
 C(x,\tau)\equiv\tav{(\varphi(x,\tau)-\varphi(0,0))^2}\,,
\end{equation}
and discuss it in various limits. Combining
(\ref{eq.densityCFcdw.exp}) and (\ref{eq.phaseCF}) we can extract
a correlation length from the relation
\begin{equation}\label{eq.corrlen}
  \xi^{-1}=\lim_{x\rightarrow \infty} \frac{p^2}{2x}C(x,0)
\end{equation}


\subsection { Disorder-free case}\label{ssec.cffree}

We start with the most simple case $u=0$. Then, the correlation
function in dimensionless units follows directly from the action
${\cal S}_0$ written in momentum space:
\begin{equation}
C_0(x,\tau)=\frac{2\pi t}{L}\sum_{k,n}\frac{1-e^{\imath(k
x+\omega_n\tau)}}{\omega_n^2+k^2}\,,
\end{equation}
with Matsubara frequencies $\omega_n=2\pi n /\lambda_T$ and
momenta $k=k_m=2\pi m/L)$.

The sums over $n$ and $k$ (i.e., $m$) can be performed
approximately for sufficiently large $x$ and $\tau$ and one
obtains~\cite{Sachdev99}
\begin{eqnarray}
 C_0(x,\tau)&&\simeq
 \frac{K}{2}\ln\Bigg(1+\label{eq.CF0}\\
 &&\left(\frac{\lambda_T }{2\pi }\right)^2
  \left[\cosh\left(\frac{2\pi x}{\lambda_T }\right)-\cos\left(\frac{2\pi
  \tau}{\lambda_T}\right)\right]\Bigg)\,.\nn
\end{eqnarray}

The behavior of this function is considered in the following
cases:

{\it (i)} At {\it zero temperature} ($\lambda_T \rightarrow
\infty$) (\ref{eq.CF0}) reduces to
\begin{equation}
C_0(x,\tau)\simeq\frac{K}{2}\ln\left(\frac{1}{2}\left[
x^2+\tau^2\right]+1\right)\,,
\end{equation}
i.e., the correlation function has a logarithmic dependency on $x$
and $\tau$ and leads to an algebraic decay of $S_1$, i.e., the
system shows QLRO.

{\it (ii)} At {\it finite temperatures} we can distinguish between
length scales smaller and larger than $\lambda_T$.

In the first case $x\ll \lambda_T$ and $\tau\ll \lambda_T$ the
$\cosh$ and $\cos$ term can be expanded to second order in the
arguments and one gets the same logarithmic function as in the
zero temperature case. In the opposite case $x\gg \lambda_T$,
which is the usual case at high temperatures, the $\cosh$ term can
be approximated by the exponential function and one finds a linear
dependency on $x$:
\begin{equation}\label{eq.CF0_highT}
 C_0(x)\approx \pi t x=Tx/ c\,\,\,\,\,\,\ \text {i.e.}
 \,\,\,\, \,\,\, \xi=\frac{2}{p^2\pi t}\equiv \xi_T,
\end{equation}
i.e., $S_1$ decays exponentially (SRO) over a characteristic
length $\xi\sim t^{-1}$. The same result is obtained for the limit
$K\to 0$ at a fixed, finite temperature.

Note, that with this result we have neglected the algebraic decay
for small $x<\lambda_T$. Therefore a better interpolation formula
for the correlation length is
$\xi\approx\frac{2}{p^2}(\xi_T+\lambda_T)$, which takes the slow
decay for small $x$ into account. In terms of the length-scale
dependent $t(l)$ this rewrites to
\begin{equation}
 t(l=\ln(\xi))=K+1\,,
\end{equation}
i.e., the correlation length is reached, if $t(l)/(1+K)$ is of
order one.

The change from QLRO on small length scales $x<\xi$ to SRO on
large length scales becomes clear if one considers the cylindric
topology of the system in space-time at finite temperatures: As
soon as one reaches length scales of the order of the perimeter of
the cylinder, which is $\lambda_T$, starting from small scales,
the system changes from two-dimensional to effectively
one-dimensional behavior.


\subsection{ Finite disorder}\label{ssec.cfdis}

If $u$ is finite the action of the system has a forward and a
backward scattering part. With the decomposition
(\ref{eq.varphide}), the phase correlation function divides into
two parts:
\begin{equation}\label{eq.phaseCFsplit}
 C(x,\tau)=C_b(x,\tau)+C_f(x)\,
\end{equation}
and has therefore always a contribution $C_f(x)\sim |x|/\xi_f$
with $\xi_f^{-1} \sim \sigma(l=\ln|x|)$, i.e., the density wave
order has always an exponentially decaying contribution and we can
define
\begin{equation}
S_1(x,\tau)\equiv f_{\rho}(x)e^{-\frac{p^2}{2}C_b(x,\tau)}\,,
\end{equation}
with
$f_{\rho}(x)=\rho_1^2\cos(pQx)e^{-\frac{p^2\pi}{4}|x|/\xi_f}$.
However, since $C_f(x)$ is not $\tau $-dependent, it will not
influence  the dynamical properties of the system. Therefore all
further remarks about phase correlations refer to $C_b(x,\tau)$
and consequently we will drop the subscript $b$ in the following.
Again we examine the $T=0$ and finite temperature cases:

{\it (i)} At {\it zero temperature} we have to distinguish between
three $K$-regimes: For $K>K_u$ the disorder becomes irrelevant
under the RG flow and we can use the zero temperature, disorder
free result for the correlation function with the pre-factor $K$
replaced by an effective quantity $K_{\textit{eff}}(l=\ln z)$ on a
length scale $z=\sqrt{x^2+\tau^2}$, defined by the flow equation
for $K$. This effective $K$ saturates on large scales at a value
$K_{\textit{eff}}(u_0)$, which may be seen in Fig.
\ref{fig.flowu}. Therefore we have QLRO in this $K$ region.

For $0<K<K_u$ we integrate the flow of $u$ until it reaches a
value of order one, starting at small $u_0$, which defines the
localization length $\xi_u$ (see section \ref{ssec.zero}), i.e.,
the correlation function behaves like $C(x,\tau)\sim |x|/\xi_u$,
i.e., we have an additional (to $C_f$) exponentially decaying
contribution to $S_1$.

\subsection{Strong disorder}\label{ssec.cfstrong}

In the last region $K=0$ we come back to the {\it strong pinning
case}, discussed in section \ref{ssec.strong} before and calculate
the pair correlation function \emph{exactly}. Taking into account
that the $h_i$'s are independent on different lattice sites, i.e.,
$\overline{h_i h_j}\propto \delta_{ij}$, the (discrete) phase
correlation function is given by
\begin{eqnarray}
 \overline{\left\langle\left(\varphi_{n+1}-\varphi_1\right)^2\right\rangle}&=&
\frac{4\pi^2}{p^2}\overline{\left\langle
\left(h_i-\frac{pQ\epsilon_i}{2\pi}-\gamma\right)^2\right\rangle}\cdot n\nn\\
&=&\frac{4\pi^2}{p^2}\overline{\left(\frac{pQ\epsilon_i}{2\pi}+\gamma-\gb{\frac{pQ\epsilon_i}{2\pi}+\gamma}\right)^2}n\,,\nn
\end{eqnarray}
where we used (\ref{eq.himin}) for the second equality. For
evaluating the disorder average in this expression, one has to
take into account the order statistics of the impurity distances
$\epsilon_i$. In the thermodynamic limit the probability density
function for the $\epsilon_i$'s can be rewritten as
\begin{equation}\label{eq.sppdf}
 p(\epsilon_i)\approx\frac{l_{\textit{imp}}^{-1}}{e}
e^{-l_{\textit{imp}}^{-1}\epsilon_i}\,, \quad
-l_{\textit{imp}}\le\epsilon_i<\infty\,.
\end{equation}
Then, the correlation function can be explicitly written as
\begin{equation}
\overline{\left\langle\left(\varphi_{n+1}-\varphi_1\right)^2\right\rangle}=
\frac{4\pi^2}{p^2}\int_{0}^{\infty}dx\,e^{-x}\left(\frac{x}{2\alpha}-\gb{\frac{x}{2\alpha}}\right)^2n\,,
\end{equation}
where we introduced the parameter $\alpha\equiv
\frac{\pi}{pQl_{\textit{imp}}}$ and substituted
$x=l_{\textit{imp}}^{-1}\epsilon_i+1$. This integral can be
evaluated exactly, which leads to the following exact expression
for the pair correlation function at zero temperature, written in
a continuum version:
\begin{equation}
C(x,\tau)=\frac{2\pi}{p\alpha}\left(1-
\frac{\alpha}{\sinh\alpha}\right)|Qx|\equiv\frac{2x}{p^2\xi},\,\,\,\,\alpha=\frac{\pi}{pQl_{\textit{imp}}}.
\label{eq.spcf}
\end{equation}
A more detailed derivation of this result is given in Appendix
\ref{app.sp}.

Finally we want give an interpolating expression for $C(x,\tau)$
from $T=0$ to high temperatures $T\gg c/(l_{\textit{imp}}p^2)$
starting with the result (\ref{eq.spcf}). In the latter case we
may neglect the discreteness of $h_i$ and hence
\begin{eqnarray}
&&\overline{\left\langle\left(\varphi_{n+1}-\varphi_1\right)^2\right\rangle}\nn\\
&&\approx \frac{4\pi^2}{p^2
l_{\textit{imp}}}\overline{\left(-\frac{\partial}{\partial
\lambda_1}\ln\left(\int dh\,e^{-\sum_i\lambda_i h^2}\right)
\right)}|x|\nn\\
&&= \frac{T}{c}|x|=\pi t |x|\,,\label{eq.cffreeT}
\end{eqnarray}
with $\lambda_i=\frac{2\pi^2 c}{T p^2
(l_{\textit{imp}}+\epsilon_i)}$.

A plausible interpolation formula is then given by
\begin{equation}    \label{eq.spcfint}
\overline{\left\langle\left(\varphi(x)-\varphi(0)\right)^2\right\rangle}
     \approx\left(2Q^2l_{\textit{imp}}\left(1-\frac{\alpha}{\sinh(\alpha)}\right)+\frac{T}{c}\right)|x|\,,
\end{equation}
and for $l_{\textit{imp}}\gg Q^{-1}$, i.e., $\alpha\ll 1$:
\begin{equation}
\overline{\left\langle\left(\varphi(x)-\varphi(0)\right)^2\right\rangle}
\approx\left(\frac{\pi^2}{3p^2}l_{\textit{imp}}^{-1}-\frac{7\pi^4}{180p^4}\frac{l_{\textit{imp}}^{-3}}{Q^2}+\frac{T}{c}\right)|x|\,.
\end{equation}
and hence the correlation length acquires the form
\begin{equation}\label{eq.xisp}
\xi_{\textit{sp}}^{-1}\approx p^2 Q^2
l_{\textit{imp}}\left(1-\frac{\alpha}{\sinh(\alpha)}\right)+\xi_T^{-1}.
\end{equation}
Note that $l_{\textit{imp}}Q\geq 1$, i.e., $\alpha\leq \pi/p$ and
$\xi_T\gg l_{imp}$. An approximate crossover to the weak pinning
limit follows by choosing $l_{\textit{imp}}\approx
L_{\textit{FL}}$.

{\it (ii)} At {\it finite temperatures} the parameter $K$
saturates at an effective value $K_{\textit{eff}}(u_0)$ on large
length scale. Therefore the correlation function for small
disorder is given by (\ref{eq.CF0}) with $K$ replaced by $K(l=\ln
z)$.

In the region ${\cal M}_u$ of the $K$-$t$ plane (see section
\ref{ssec.crossover}), $u$ still increases and we can find an
effective correlation length by comparing the length scales on
which $u(l)$ or $t(l)/(1+K(l)$ become of order one. Then, the
correlation length is the smaller length of these two.

\subsection {Burgers equation}\label{ssec.burgers}


For $K=0$, high temperatures but weak disorder we adopt an
alternative method by mapping the (classical) one-dimensional
problem onto the Burgers equation with
noise~\cite{HuseHenleyFi85}. With this approach one can derive an
effective correlation length given by
\begin{equation}\label{eq.xiburgers}
  \xi_B^{-1}\approx \xi_T^{-1}\left(1+\frac{1}{2}(
  \frac{\xi_T}{2L_{FL}})^3\right)
\end{equation}
where $\xi_T \ll L_{FL}$, which changes the prefactor of the free
correlation function at high temperatures (\ref{eq.CF0_highT}).
The full calculation of this result can be found in Appendix
\ref{app.weak}.

\begin{table}[htb]
\caption{\label{tab.cl} Overview of the dimensionless correlation
lengths.}
\begin{ruledtabular}
\begin{tabular}{clc}
length& description & eq.\\
\hline
$\xi_B$ & weak pinning/high temp. length &(\ref{eq.xiburgers})\\
$\xi_f$ & forward scattering length &(\ref{eq.phaseCFsplit})\\
$\xi_{\textit{sp}}$ & strong pinning length &(\ref{eq.xisp})\\
$\xi_T$ & high temp./disorder free length & (\ref{eq.CF0_highT})\\
$\xi_u$ & disorder localization length & (\ref{eq.xiu})\\
$\xi_w$ & lattice pot. correlation length &sec. \ref{sec.lp}\\
\end{tabular}
\end{ruledtabular}
\end{table}


\section{Superfluids}\label{sec.sf}

Next we consider the application of the results obtained so far to
a {\it one-dimensional Bose fluid}.  Its density operator is given
by eq. (1) if we identify $Qf/\pi=\rho_0=\rho_1$ ($p=2$):
\begin{equation}\label{eq.sfdensity}
 \rho_{\textit{SF}}=\frac{f}{\pi}\partial_x\varphi+\rho_0(1+\cos(2(\varphi+Qx)))+\ldots
\end{equation}
$\partial_x\varphi$ is conjugate to the phase $\theta$ of the Bose
field operator~\cite{Haldane}. Keeping our definitions of $K$, $t$
and $u$, $v$ denotes now the phase velocity of the sound waves
with $v=\sqrt{\kappa/(\rho_0m)}$ and the elastic constant is
$c=\kappa/(\pi\rho_0)^2$, where $\kappa$ is the compressibility
per unit length (see also table \ref{tab.nguide}).

With the replacements
\begin{eqnarray}
 K&\rightarrow& K^{-1}\nn\\
 t&\rightarrow& t/K^2\nn\\
 p&=&2\,,\nn
\end{eqnarray}
(\ref{eq.S^n/hbar}) describes the action of the 1D-superfluid in a
random potential. The correlation functions for the superfluid can
be obtained correspondingly from this replacements. To avoid
confusions we write down the full action in this case explicitly:

\begin{widetext}

\begin{eqnarray}
&&\frac{{\cal
S}_{\textit{SF}}}{\hbar}=\frac{K}{2\pi}\sum_{\alpha,\beta}
\int\limits_{0}^{L}dx\int\limits_0^{K/t}d\tau\label{eq.Ssf}\\
&&\Bigg\{
\Big[(\partial_x\varphi_{\alpha})^2+(\partial_{\tau}\varphi_{\alpha})^2\Big]
\delta_{\alpha\beta} -
\frac{K}{2}\int\limits_0^{K/t}d\tau^{\prime}\Big[u^2\cos{2
\Big(\varphi_{\alpha}(x,\tau)-\varphi_{\beta}(x,\tau^{\prime})\Big)}
+\sigma\partial_x\varphi_{\alpha}(x,\tau)\partial_x\varphi_{\beta}(x,\tau^{\prime})\Big]\Bigg\}\,.\nn
\end{eqnarray}
\end{widetext}

Hence the RG-equations follow from (\ref{eq.dt/dl}) to
(\ref{eq.ds/dl}) with the above given replacements:
\begin{eqnarray}
\frac{dt}{dl}
&=&\Big[1+\frac{16u^2}{K^2}B_0\left(4/K,\frac{K}{2t}\right)
\coth \frac{K}{2t}\Big]t\,,\nn\\
\frac{dK}{dl} & = & \frac{8u^2}{K}B_0\left(4/K,\frac{K}{2t}\right)
\coth \frac{K}{2t}\nn,\\
\frac{du^2}{dl} & = & \Big[3  -
\frac{2}{K}\coth{\frac{K}{2t}}\Big]u^2\,,\nn\\
\frac{d\sigma}{dl}&=&\sigma\,.\nn
\end{eqnarray}

I.e., the transition between the superfluid and the localized
phase occurs at $K_u^*=2/3$ [\onlinecite{GiaSchulz87}]. Thermal
fluctuations again suppress the disorder and destroy the the
superfluid localization transition in 1D.

\section{Lattice potential}\label{sec.lp}

If the wave length $\lambda$ of the CDW modulation is commensurate
with the period $a$ ($=\pi$, due to dimensionless units) of the
underlying lattice such that $n\lambda=qa$ with integers $n$ and
$q$, the Umklapp term $-2\pi (w/K)\cos(q\varphi)$ appears in the
Hamiltonian~\cite{Gruener}. Therefore we switch on the lattice
potential $w\neq 0$ now. In this section we consider the case
$u=0$ which leads to the Sine-Gordon type model:
\begin{equation}
 \frac{{\cal S}_{\textit{LP}}}{\hbar}=\int\limits_0^L dx\int\limits_0^{K/t} d\tau\frac{1}{2\pi
 K}\left\{(\partial_x\varphi)^2+(\partial_{\tau}\varphi)^2\right\}-\frac{w}{K}\cos(q\varphi)
\end{equation}

The model has $q$ degenerate classical ground states given by
$\varphi_m=2\pi m/q$ with $m=0,...,q-1$.    Performing a
calculation analogous to the one above (but with $u=0$) the
RG-flow equations read
\begin{subequations}
 \begin{eqnarray}
    \frac{dK}{dl} & = & \frac{\pi}{2}q^4w^2
   B_2\Big(q^2 K,\frac{K}{2t}\Big)\coth{\frac{K}{2t}},
\label{eq.dK/dl.sg}\\
\frac{dt}{dl} & = & \left[1+\frac{\pi}{2}q^4w^2
   B_1\Big(q^2 K,\frac{K}{2t}\Big)\coth{\frac{K}{2t}}\right]t,
   \label{eq.dt/dl.sg}\\
   \frac{dw}{dl} & = & \left[2-\frac{q^2}{4}K\coth{\frac{K}{2t}}\right]w,
   \label{eq.dw/dl.sg}
\end{eqnarray}
\end{subequations}
where $B_{1,2}$ are given in (\ref{eq.B}) with
\begin{eqnarray}
g_1&=&2x^2\cos{x}\,,\nn\\
g_2&=&(x^2+\tau^2)\cos{x}\,.\nn
\end{eqnarray}
Plots of the functions $B_1$ and $B_2$ can be found at the end of
Appendix \ref{app.RG}.

At zero temperature (\ref{eq.dK/dl.sg}) and (\ref{eq.dw/dl.sg})
reduce to
\begin{subequations}
 \begin{eqnarray}
\frac{dK}{dl} & = & \frac{\pi}{2}q^4w^2
   B_2\Big(q^2K,\infty\Big),
\label{eq.dK/dl.sg0}\\
   \frac{dw}{dl} & = & \left[2-\frac{q^2}{4}K\right]w,
   \label{eq.dw/dl.sg0}
\end{eqnarray}
\end{subequations}
and we find, that for $u=0$ the lattice potential becomes relevant
(i.e., $w$ grows) for $K<K_w$, where $K_w$ is implicitly defined
by
\begin{equation}\label{eq.Kw}
 w^2(K_w)=\frac{K_w^{*2}}{2\pi q^2\tilde\eta}\left(\frac{K_w}{K_w^*}-1\right)^2\,,
\end{equation}
which follows from
\begin{equation}
\frac{dw}{dK}=-\frac{4}{q^4\pi\tilde\eta
w}\left(1-\frac{K}{K_w^*}\right)\,,
\end{equation}
where we used (\ref{eq.dK/dl.sg0}) and (\ref{eq.dw/dl.sg0}) and
the initial condition $w(K_w^*\equiv 8/q^2)=w_0=0$;
$\tilde\eta=-B_2\Big(q^2 K_w^*,\infty\Big)$ ($\approx 0.4$, for
$q=1$).

In this region the periodic potential stabilizes  true long-range
order of the CDW: the phase is everywhere close to one of the $q$
classical ground states $\varphi_m$. The {\it depinning transition
from the lattice} for $K\rightarrow K_w-0$ is again of KT type.
The correlation length in the low-$K$ ordered phase $\xi_{w}$ is
defined by $w(\ln\xi_{w})\approx 1$ and diverges at $K_w-0$
[\onlinecite{FisherGrinstein}]. This can be seen by considerations
analogous to  the disordered case. Defining
\begin{eqnarray}
\gamma&=& 2\frac{K}{K_w^*}-2\,,\nn\\
\chi^2&=&\frac{\pi}{8}q^6\tilde\eta w^2\nn
\end{eqnarray}
(note that $\tilde\eta>0$) leads for $|\gamma|\ll 1$, i.e., close
to $K_w^*$, to the KT equations (\ref{eq.KTchi}) and
(\ref{eq.KTgamma}).

\begin{figure}[htb]
\includegraphics[width=0.6\linewidth]{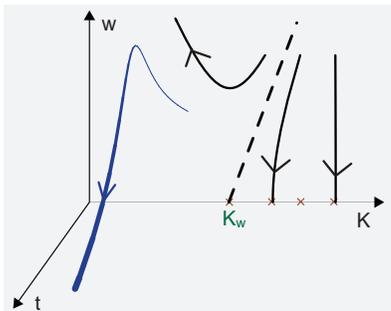}
\caption{Typical flow diagram for the disorder free model
 in the three dimensional parameter space of $K$, $w$ and $t$.
 $w$ denotes the strength of the commensurate lattice potential}.
 \label{fig.floww}
\end{figure}

At finite temperatures we find a similar scenario as in the case
where we considered the influence of the disorder: $w$ first
increases in a $K$--$t$ region given by ${\cal
M}_w\equiv\left\{(K,t)|K^{\ast}_w\geq K\coth{\frac{K}{2t}}\geq
0\right\}$, i.e., when the r.h.s. of (\ref{eq.dw/dl.sg}) is
positive, but then decreases and flows into the region of large
$t$ and small $w$. Thus the periodic potential becomes irrelevant
at finite temperatures. This can be understood as follows: at
finite $t$ the 1D quantum sine-Gordon model can be mapped on the
Coulomb gas model on a torus of perimeter $\lambda_T$ since
periodic boundary conditions apply now in the $\tau$-direction.
Whereas the entropy of two opposite charges increases for
separation $L\gg \lambda_T$ as $\ln (L\lambda_T)$, their action
increases linearly with $L$. Thus, the charges remain bound.  The
one-dimensional Coulomb gas has indeed only an insulating phase
~\cite{Lenard}.

\section{Conclusion}\label{sec.concl}

To conclude we have shown that in one-dimensional charge and spin
density waves, Luttinger liquids and suprafluids, quantum phase
transitions between a disordered (or locked-in) phase and an
asymptotically free phase at zero temperature are destroyed by
thermal fluctuations leaving behind a rich crossover behavior.
This was demonstrated by using a {full finite temperature}
renormalization group (RG) calculation. The crossover regions were
characterized by the behavior of the phase pair correlation
functions. For vanishing quantum fluctuations our calculation was
amended by an {exact} solution in the case of strong disorder and
by a mapping onto the {Burgers equation with noise} in the case of
weak disorder, respectively. Both methods gave an exponential
decay of density correlations.

We have also briefly discussed, that the Coulomb interaction may
destroy the unpinning (localization) transition.

The finite temperature calculation used in the present paper is
also suited for treating the low frequency low temperature
behavior of dynamical properties which may depend crucially on the
ratio $\omega/T$. This as well as the discussion of the influence
of quantum phase slips~\cite{GlaNa} will be postponed to a
forthcoming publication.

The combined effect of disorder  and the lattice potential on the
zero temperature phase diagram is still controversely
discussed~\cite{Shankar90,Orignac} and cannot be explained by the
RG-results presented in this paper, since both pertubations become
relevant for small $K$. However, this problem is beyond the scope
of the present work and may be discussed in a future publication.

\section{Acknowledgement}
The authors thank A. Altland, S. Brasovskii, T. Emig, S.
Korshunov, B. Rosenow, S. Scheidl and V. Vinokur for useful
discussions. They acknowledge financial support by Deutsche
Forschungsgemeinschaft through Sonderforschungsbereich 608. AG
further acknowledges financial support by Deutscher Akademischer
Austauschdienst.

\appendix

\section{renormalization of the disorder}\label{app.RG}

We present a short overview of the used finite temperature
anisotropic renormalization group method in the case of the
replicated disorder term. Starting point is the action
(\ref{eq.S^n/hbar}).

The phase field $\varphi(x,\tau)=\frac{t}{K
L\Lambda}\sum_{\omega_n}\sum_{|k|\le
1}e^{\imath(\omega_n\tau+kx)}\varphi_{k,\omega_n}$ with the
Matsubara frequencies $\omega_n=2\pi n t/K$ and $k=2\pi
m/(L\Lambda)$ (note that rescaled coordinates are used) is split
in a slow ($|k|<b^{-1}$) and a fast mode part ($b^{-1}\le |k|\le
1$), where $b=e^{-dl}$ is a rescaling parameter of order $1$.
Notice, that the $\varphi_{\lessgtr}$ still have all Matsubara
Fourier components.

In order to find the RG-corrections of the other parameters in the
model, we follow {\it Wilson}~\cite{Wilson73} and expand
$\cav{e^{-{\cal S}_u^{(n)}/\hbar}-1}_{0,>}$ in small $(u/K)^2$,
with
\begin{equation}
\frac{{\cal S}_u^{(n)}}{\hbar}=-\frac{u^2}{4\pi
K^2}\sum\limits_{\alpha,\beta}\iint d\tau d\tau^{\prime}\,\int dx
\, R[\varphi_{\alpha}(x,\tau)-\varphi_{\beta}(x,\tau^{\prime})]
\end{equation}
where $R[f]\equiv \cos(p f)$. $\cav{\ldots}_{0,>}$ denotes the
cumulative or connected average over the fast modes in the
''momentum stripes'' with the free gaussian model. The correction
in first order is given by
\begin{equation}
\frac{{\cal S}_{u,1}^{(n)}}{\hbar}=\cav{\frac{{\cal
S}_u^{(n)}}{\hbar}}_{0,>}\,.
\end{equation}
For calculating the cumulative average of the functional $R$, $R$
is expanded in small $\Delta\varphi_{>}\equiv
\varphi_{\alpha,>}(x,\tau)-\varphi_{\beta,>}(x,\tau^{\prime})$,
e.g.

\begin{widetext}

\begin{eqnarray}
\cav{R[\Delta\varphi]}_{0,>}=-p^2\big(R[\Delta\varphi_{<}]\tav{\varphi_{\alpha,>}^2}_{0,>}-
R[\Delta\varphi_{<}]\tav{\varphi_{\alpha,>}(x,\tau)\varphi_{\beta,>}(x,\tau^{\prime})}_{0,>}\big)+{\cal
O}(\Delta\varphi_{>}^4)\,. \label{eq.Rexp}
\end{eqnarray}

The first term in (\ref{eq.Rexp}) gives a correction to the
disorder parameter $u$ and the second term a correction to $K$.
The free thermal average over the fast modes can be evaluated with
the free propagator $(k^2+\omega_n^2)^{-1}$ and using the formula
\begin{equation*}
  \sum\limits_{n=-\infty}^{\infty}\frac{\cos(n x)}{n^2+a^2}=
   \frac{\pi}{|a|}\frac{\cosh((\pi-x)a)}{\sinh(\pi |a|)}\;,\; 0\le x\le 2\pi
\end{equation*}
to treat the {\it sum} over the Matsubara frequencies yielding
\begin{equation}
\tav{\varphi_{\alpha,>}(x,\tau)\varphi_{\beta,>}(x,\tau^{\prime})}_{0,>}=
 \frac{K}{2}\frac{\cosh(K/2t-|\Delta\tau|)}{\sinh(K/2t)}\delta_{\alpha,\beta}\ln
 b\,,\label{eq.A3}
\end{equation}
with $\Delta\tau\equiv \tau-\tau^{\prime}$.

In order to find a good gaussian approximation for $R$, we perform
a variational calculation for a Sine--Gordon model:
\begin{equation}
{\cal H}_{\textit{SG}}={\cal H}_0+{\cal H}_1\equiv{\cal
H}_0-\mu\int d^dr\,\cos(\varphi(\vec{r}))\,,
\end{equation}
where ${\cal H}_0$ is the gaussian part. We approximate ${\cal
H}_1$ by
\begin{equation}
\tilde {\cal H}_1=\int
d^dr\,\frac{\kappa(\vec{r})}{2}\varphi^2(\vec{r})\,,
\end{equation}
and define $\tilde {\cal H}\equiv {\cal H}_0+\tilde {\cal H}_1$.
To find the optimal function $\kappa(\vec{r})$, the {\it
variational free energy}~\cite{Fvar} ${\cal F}_{\textit{var}}$,
with
\begin{equation}
{\cal F}_{\textit{SG}}\leq {\cal F}_{\textit{var}}\equiv \tilde
{\cal F}+\tav{{\cal H}_{\textit{SG}}-\tilde {\cal H}}_{\tilde
{\cal H}}\,
\end{equation}
is minimized with respect to $\kappa$:

\begin{eqnarray}
0=\frac{\delta {\cal F}_{\textit{var}}}{\delta\kappa(\vec{\tilde
r})}&=&\beta\tav{{\cal H}_{\textit{SG}}-\tilde {\cal H}}_{\tilde
{\cal H}}\tav{\frac{\varphi^2(\vec{\tilde r})}{2}}_{\tilde {\cal
H}}-\beta\tav{({\cal H}_{\textit{SG}}-\tilde{\cal
H})\frac{\varphi^2(\vec{\tilde r})}{2}}_{\tilde {\cal H}}\nn\\
&=&\frac{\beta}{2}\int
d^dr\,\tav{\kappa(\vec{r})-\mu\cos(\varphi(\vec{r}))}_{\tilde
{\cal H}}\cav{\varphi^2(\vec{r})\varphi^2(\vec{\tilde r})}_{\tilde
{\cal H}}\label{eq.A9}\,.
\end{eqnarray}
For the last equality we took into account that the averages are
gaussian such that we could apply the {\it Wick theorem}. From
(\ref{eq.A9}) we finally get
\begin{equation}
\kappa(\vec{r})=\mu\tav{\cos(\varphi(\vec{r}))}_{\tilde {\cal
H}}=\mu \exp\left(-\frac{1}{2}\tav{\varphi^2(\vec{r})}_{\tilde
{\cal H}}\right)\label{eq.A10}\,.
\end{equation}

For small disorder (\ref{eq.A10}) yields for
$R[\Delta\varphi_{<}]$ the approximate expression
\begin{equation}\label{eq.A4}
 R[\Delta\varphi_{<}]\simeq
 -\frac{p^2}{2}(\Delta\varphi_{<})^2e^{-\frac{p^2}{2}\tav{(\Delta\varphi_{<})^2}_{0,<}}\,.
\end{equation}
The same result can be obtained in terms of an operator product
expansion~\cite{Knops80} of $R[f]$. In order to get a RG
correction to $K$ a gradient expansion of $(\Delta\varphi_{<})^2$
in (\ref{eq.A4}) in small $\Delta\tau$ is performed, which is
justified by the exponential decay of the correlation function
(\ref{eq.A3}) on the integration interval such that higher orders
in $\Delta\tau$ do not contribute to the RG correction:
$(\Delta\varphi_{<})^2\approx
(\partial_{T}\varphi_{<}(x,T)\Delta\tau)^2$ with
$T=(\tau+\tau^{\prime})/2$.

The pair correlation function in the argument of the exponential
function in (\ref{eq.A4}) can be approximated by the expression
already shown in (\ref{eq.CF0}), i.e.,
\begin{equation}\label{eq.CF0_2}
 \tav{(\varphi(x,\tau)-\varphi(0,0))^2}_0\simeq
 \frac{K}{2}\ln\left(1+\left(\frac{K}{2\pi t}\right)^2
  \left[\cosh\left(\frac{2\pi t x}{K}\right)-\cos\left(\frac{2\pi t
  \tau}{K}\right)\right]\right)\,.
\end{equation}

\end{widetext}

After integration of the fast modes of $\varphi$, one rescales the
system to maintain the fluctuation strength and the spatial
density of the degrees of freedom ($l=\ln b$):
\begin{eqnarray}
 x\;&\rightarrow&\; x^{\prime}=xb^{-1}\nn\\
 \tau\;&\rightarrow&\; \tau^{\prime}=\tau b^{-z}\;,\quad T\;\rightarrow\; T^{\prime}=Tb^{z}\nn\\
 \varphi\;&\rightarrow&\; \varphi^{\prime}=\varphi b^{-\zeta}\nn
\end{eqnarray}
which leads to rescaled parameters. For our model these are given
by (here in $d$ dimensions)
\begin{subequations}
\begin{eqnarray}
 c^{\prime}&=&cb^{d+z-2+2\zeta}\,,\\
 v^{\prime}&=&vb^{z-1}\,,\\
 K^{\prime}&=&Kb^{1-d-2\zeta}\,,\label{eq.rescaleK}\\
 t^{\prime}&=&tb^{2-d-2\zeta}\,,\\
 u^{\prime}&=&ub^{2-d/2}\,,\label{eq.rescaleu}\\
 \sigma^{\prime}&=&\sigma b^{2-d}\,,\label{eq.rescales}\\
 w^{\prime}&=&wb^{2d}\,.
\end{eqnarray}
\end{subequations}
Due to the invariance of the system under a phase shift of
$2n\pi$, $\zeta$ is zero for symmetry reasons.

The RG contribution to the flow equation for $u^2$  follows from
the first term of (\ref{eq.Rexp}) and (\ref{eq.A3}) with
$\Delta\tau=0$:
\begin{equation}
{u^{\prime}}^2=u^2\left(1-\frac{p^2 K}{2}\coth\frac{K}{2 t}\ln
b\right)\,.
\end{equation}
Together with (\ref{eq.rescaleu}) ($d=1$) one gets
\begin{equation}
\frac{{u^{\prime}}^2-u^2}{dl}=\left(3-\frac{p^2
K}{2}\coth\frac{K}{2 t}\right)u^2\,.
\end{equation}

The RG correction to $K$ follows from the second term of
(\ref{eq.Rexp}) with (\ref{eq.A3}) and (\ref{eq.A4}):
\begin{eqnarray}
K^{\prime}=K\Bigg(1-&&\frac{p^4}{2}u^2\coth\frac{K}{2t}\int\limits_0^{K/(2t)}
d\tau\,\tau^2\frac{\cosh\left(\frac{K}{2t}-\tau\right)}{\cosh\frac{K}{2t}}\times\nn\\
&&\qquad
e^{-\frac{p^2}{2}\tav{(\varphi(0,\tau)-\varphi(0,0))^2}_0\ln
b}\Bigg)\,,
\end{eqnarray}
and the flow equation (for $d=1+\epsilon$) follows from
(\ref{eq.rescaleK}):
\begin{equation}\label{eq.flowK}
\frac{K^{\prime}-K}{dl}=\left[-\epsilon-\frac{p^4}{2}u^2\coth\frac{K}{2t}B_0\left(p^2K,\frac{K}{2t}\right)\right]K
\end{equation}
with $B_0$ given in (\ref{eq.B}), for which we used
(\ref{eq.CF0_2}).

\begin{figure}[htb]
\includegraphics[width=0.8\linewidth]{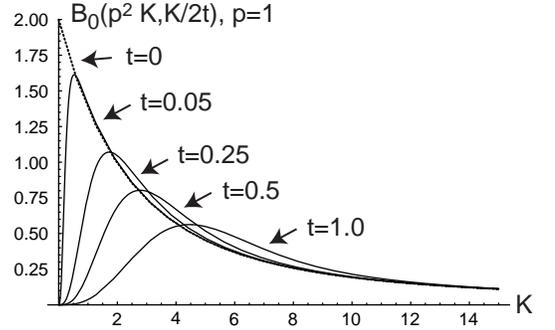}
\caption{Function $B_0(p^2 K,\frac{K}{2t})$ plotted with respect
to $K$ for different temperatures. }\label{figB0}
\end{figure}
The function $B_0$ which appears in this flow equation is plotted
in Fig. \ref{figB0}.

\begin{figure}[htb]
\includegraphics[width=0.8\linewidth]{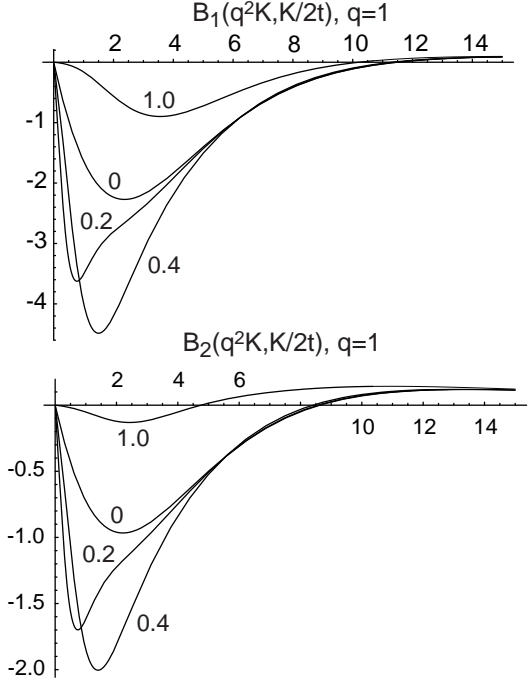}
\caption{Function $B_1(q^2K,\frac{K}{2t})$ and
$B_2(q^2K,\frac{K}{2t})$ plotted with respect to $K$ for different
temperatures, written next to the graphs, and $q=1$.
}\label{figB1B2}
\end{figure}

For completeness we also plot the functions $B_1$ and $B_2$ in the
relevant $K$-region for the lattice unpinning transition [see eqs.
(\ref{eq.dK/dl.sg}) and (\ref{eq.dt/dl.sg})]. Note, that for
evaluating these functions at zero temperature, one has to execute
the integrals at finite temperature first and then take the limit
$t\to 0$.

\section{strong pinning}\label{app.sp}

To calculate the phase correlation function in the strong pinning
limit, it is necessary to study the order statistics of the
impurity distances $\epsilon_i=(x_i-x_{i-1})-l_{\textit{imp}}$.
Following David~\cite{David}, we obtain for the probability
density function (pdf) of the $\epsilon_i$'s in the case of
uniformly distributed impurity positions $0\le x_i\le L$ :
$p(\epsilon_i)=l^{-1}_{\textit{imp}}(1-\frac{1}{L}(\epsilon_i+l_{\textit{imp}}))^{N-1}$
with $-l_{\textit{imp}}\le\epsilon_i\le L-l_{\textit{imp}}$. In
the thermodynamic limit the pdf can be rewritten as
\begin{equation}
 p(\epsilon_i)\approx\frac{l_{\textit{imp}}^{-1}}{e}
e^{-l_{\textit{imp}}^{-1}\epsilon_i}\,, \quad
-l_{\textit{imp}}\le\epsilon_i<\infty\,.
 \label{eq.pdfe}
\end{equation}

With this, one can calculate the $n$-th moment
$\overline{\epsilon_i^n}$ (for $n>1$, $\overline{\epsilon_i}=0$)
as follows:
\begin{eqnarray}
 \overline{\epsilon_i^n}&=&\int_{-l_{\textit{imp}}}^{\infty}
\frac{l_{\textit{imp}}^{-1}}{e}e^{-l_{\textit{imp}}^{-1}\epsilon_i}
\epsilon_i^n\,d\epsilon_i\nonumber\\
 &=&\frac{l_{\textit{imp}}^{-1}}{e}(-1)^n\left.
\frac{\partial^n}{\partial \lambda^n}\right|_{\lambda=l_{\textit{imp}}^{-1}}
 \frac{e^{\lambda l_{\textit{imp}}}}{\lambda}\,.
\end{eqnarray}

Using $\left.\frac{\partial^n}{\partial x^n}\right|_{x=1}\frac{e^{x-1}}{x}=
(-1)^n n!\left(\sum_{k=1}^{n}\frac{(-1)^k}{k!}+1\right)$ yields:
$\overline{\epsilon_i^n}=\frac{n!}{c^n}\sum_{k=2}^{n}\frac{(-1)^k}{k!}$
and for the correlator
$\overline{\epsilon_i\epsilon_j}=l_{\textit{imp}}^{2}\delta_{ij}$.

With this results we can derive the pair correlation function
(\ref{eq.spcf}). Therefore we calculate the discrete version
$\overline{\left\langle\left(\varphi_n-\varphi_1\right)^2\right\rangle}$
in the limit $T\rightarrow 0$. With the definitions given below
eq. (\ref{eq.strongham}), we can rewrite:
\begin{equation}
 \left(\varphi_n-\varphi_1\right)^2
    =\frac{4\pi^2}{p^2}\left(\sum_{i=1}^{n-1}\left(h_i-\frac{pQ\epsilon_i}{2\pi}-\gamma\right)\right)^2\,.
\end{equation}

Using $\overline{\epsilon_i}=\overline{h_i}=0$ and
$\overline{h_i h_j}\propto \delta_{ij}$ leads to
\begin{equation}
 \overline{\left\langle\left(\varphi_{n+1}-\varphi_1\right)^2\right\rangle}=
\frac{4\pi^2}{p^2}\underbrace{\overline{\left\langle
\left(h_i-\frac{pQ\epsilon_i}{2\pi}-\gamma\right)^2\right\rangle}}_{\equiv
\mathrm{\tilde C}}\cdot n\,.
\end{equation}

Because only the value $\gb{\frac{pQ\epsilon_i}{2\pi}+\gamma}$ for
$h_i$ is taken into account for evaluation of the thermal average,
we get
\begin{eqnarray}
   &&\mathrm{\tilde C}=
   \overline{\left(\frac{pQ\epsilon_i}{2\pi}+\gamma-\gb{\frac{pQ\epsilon_i}{2\pi}+\gamma}\right)^2}\nonumber\\
   &&=\int_{-l_{\textit{imp}}}^{\infty}d\epsilon_i\,\frac{l_{\textit{imp}}^{-1}}{e}
   e^{-l_{\textit{imp}}^{-1}\epsilon_i}\left(\frac{pQ\epsilon_i}{2\pi}+\gamma-\gb{\frac{pQ\epsilon_i}{2\pi}+\gamma}\right)^2\,.\nn
\end{eqnarray}
If we substitute $x=l_{\textit{imp}}^{-1}\epsilon_i+1$ and take
into account that $\gb{x+n}=\gb{x}+n$ for $n\in \mathbb{Z}$, we
get
\begin{equation}
 \mathrm{\tilde C}=
  \int_{0}^{\infty}dx\,e^{-x}\left(\frac{x}{2\alpha}-\gb{\frac{x}{2\alpha}}\right)^2
\end{equation}
with the parameter $\alpha=\frac{\pi}{pQ l_{\textit{imp}}}$. Now
the quadratic term in the integral is expanded, which leads to the
following three (converging) integrals
\begin{eqnarray}
 I_1&\equiv &\frac{1}{4\alpha^2}\int_{0}^{\infty}dx\,e^{-x}x^2=\frac{1}{2\alpha^2}\,,\nonumber\\
 I_2&\equiv & -\frac{1}{\alpha}\int_{0}^{\infty}dx\,e^{-x}x\gb{\frac{x}{2\alpha}}\,,\nonumber\\
    & =& -\frac{1}{\alpha}\sum_{k=1}^{\infty} \int\limits_{(2k-1)\alpha}^{(2k+1)\alpha}dx\,e^{-x}x\gb{\frac{x}{2\alpha}}\\
 I_3&\equiv
 &\int_{0}^{\infty}dx\,e^{-x}\gb{\frac{x}{2\alpha}}^2\,,\nn\\
     &=&\sum_{k=1}^{\infty} \int\limits_{(2k-1)\alpha}^{(2k+1)\alpha}dx\,e^{-x}\gb{\frac{x}{2\alpha}}^2\,, \nonumber
\end{eqnarray}
with $\mathrm{\tilde C}=I_1+I_2+I_3$. For $(2k-1)\alpha\le x\le
(2k+1)\alpha$, $k\in\mathbb{Z}$, $\gb{\frac{x}{2\alpha}}= k$, such
that the {\it Gaussian brackets} in the finite integrals in $I_2$
and $I_3$ can be replaced by $k$ or $k^2$, respectively:
\begin{eqnarray}
 I_2&=&-\frac{1}{\alpha}\sum_{k=1}^{\infty} k\int\limits_{(2k-1)\alpha}^{(2k+1)\alpha}dx\,e^{-x}x\,,\nonumber\\
 I_3&=&\sum_{k=1}^{\infty}k^2
 \int\limits_{(2k-1)\alpha}^{(2k+1)\alpha}dx\,e^{-x}\,.
\end{eqnarray}
The values of these two simple integrals are
\begin{eqnarray}
 \int\limits_{(2k-1)\alpha}^{(2k+1)\alpha}dx e^{-x}x&=&2e^{-2k\alpha}((1+2k\alpha)\sinh(\alpha)-\nn\\
&&\qquad\alpha\cosh(\alpha))\,,\nonumber\\
 \int\limits_{(2k-1)\alpha}^{(2k+1)\alpha}dx e^{-x}&=&2e^{-2k\alpha}\sinh(\alpha)\,.\nn
\end{eqnarray}

The remaining sums in $I_2$ and $I_3$ are only derivatives of the
geometric series which can be easily evaluated.

The result is
\begin{eqnarray}\label{eq.spcfdis}
 \mathrm{\tilde C}&=&\frac{1}{2\alpha}\left(\frac{1}{\alpha}-\frac{1}{\sinh(\alpha)}\right)\\
 &=&\frac{1}{12}-\frac{7}{720}\alpha^2+\frac{31}{30240}\alpha^4+\mathcal{O}(\alpha^6)\,,\nn
\end{eqnarray}
where the expansion is useful only if $\alpha$ is small, i.e., $Q
l_{\textit{imp}}\gg 1$.
If we would have neglected the order
statistics of the impurity distances we would get only the leading
constant: $\mathrm{\tilde C}=1/12$.

(\ref{eq.spcfdis}) yields the presented expression for the pair
correlation function (\ref{eq.spcf}).

\section{correlation length in the classical
region at finite temperature}\label{app.weak}

In the weak pinning limit, $1\ll l_{\textit{imp}}Q \ll
c/(U_{\textit{imp}}\rho_1)$ or ${L_{\textit{FL}}}\gg
l_{\textit{imp}}$, the classical Hamiltonian can be rewritten to a
{\it random-field XY-model}

\begin{eqnarray}
 {\cal H}_{\textit{class}}(L)&=&
 \int\limits_{0}^{L}dx\,\Bigg\{\frac{c}{2}\left(\partial_x\varphi(x)-
 \tilde\sigma\right)^2+\nn\\
 &&\frac{U_{\textit{imp}}\rho_1}{l_{\textit{imp}}}
\cos[p(\varphi(x)-\alpha(x))]\Bigg\}\label{eq.weakH}
\end{eqnarray}

where $\alpha(x)$ is a random phase with zero average and
\begin{equation}\label{eq.alphaav}
 \dav{e^{\imath p\big(\alpha(x)-\alpha(x^{\prime})\big)}}=l_{\textit{imp}}\delta(x-x^{\prime}).
\end{equation}
In the following we consider only the backward scattering term of
the correlation function $C_b$ and therefore neglect the forward
scattering term $\tilde\sigma=U_{\textit{imp}}f/(c\pi
l_{\textit{imp}})$.

The goal is now to find an effective temperature, which replaces
$T$ in the correlation function for the free case
(\ref{eq.cffreeT}). We start with a {\it Burgers}-like equation,
which one gets after a {\it Cole-Hopf transformation} from the
transfer matrix equation, for the {\it restricted} free energy
${\cal F}(x,\varphi)=-T\ln {\cal Z}(x,\varphi)$ with the partition
function
\begin{equation*}
{\cal
Z}(x,\varphi)\equiv\int\limits_{\varphi(0)}^{\varphi(x)=\varphi}{\cal
D}\varphi\,e^{-H(x)/T}\,.
\end{equation*}
The equation reads
\begin{equation}\label{eq.burgers}
\frac{\partial {\cal F}}{\partial x}=\frac{T}{2c}\frac{\partial^2
{\cal F}}{\partial \varphi^2}-\frac{1}{2c}\left(\frac{\partial
{\cal F}}{\partial
\varphi}\right)^2+\underbrace{\frac{U_{\textit{imp}}\rho_1}{l_{\textit{imp}}}
\cos[p(\varphi(x)-\alpha(x))]}_{U(x,\varphi)}\,.
\end{equation}
Using the Fourier transform ${\cal
F}(x,\varphi)=\int\frac{dk\,d\omega}{(2\pi)^2}e^{\imath(\omega\varphi-kx)}{\cal
F}(k,\omega)$ (analogous for $U(x,\varphi)$), (\ref{eq.burgers})
is rewritten as
\begin{eqnarray}
&-\imath k{\cal F}(k,\omega)=-\frac{T\omega^2}{2c}{\cal
F}(k,\omega)+U(k,\omega)+&\label{eq.burgersft}\\
&\frac{1}{2c}\int\frac{dk^{\prime}\,d\omega^{\prime}}{(2\pi)^2}\omega^{\prime}(\omega-\omega^{\prime}){\cal
F}(k-k^{\prime},\omega-\omega^{\prime}){\cal
F}(k^{\prime},\omega^{\prime})&\,,\nn
\end{eqnarray}
with
\begin{eqnarray}
U(k,\omega)&=&\frac{\pi
U_{\textit{imp}}\rho_1}{l_{\textit{imp}}}\bigg\{
h_{+}(k)\delta(\omega-p)\nn\\
&&\qquad\qquad+h_{-}(k)\delta(\omega+p)\bigg\}\,,\label{eq.UKw}\\
h_{\pm}(k)&\equiv&\int dx\, e^{\imath[kx\mp p\alpha(x)]}\,.\nn
\end{eqnarray}

Introduction the dimensionless quantities
\begin{eqnarray}
g_0(k,\omega)&=&\frac{1}{\pi t\omega^2/2-\imath k/\Lambda}\,,\\
\epsilon&=&\frac{U_{\textit{imp}}\rho_1}{l_{\textit{imp}}\Lambda^2
c}\,,\\
u(k,\omega)&=&(\epsilon\Lambda c)^{-1}U(k,\omega)\,,
\end{eqnarray}
and setting ${\cal F}(k,\omega)\equiv c\epsilon
g(k,\omega)u(k,\omega)$ we obtain the following, self-consistent
equation for the Green's function $g(k,\omega)$:
\begin{widetext}
\begin{equation}\label{eq.burgerg}
g(k,\omega)u(k,\omega)=g_0(k,\omega)u(k,\omega)+\frac{\epsilon}{2}
g_0(k,\omega)\int\frac{dk^{\prime}\,d\omega^{\prime}}{\Lambda(2\pi)^2}
\omega^{\prime}(\omega-\omega^{\prime})g(k-k^{\prime},\omega-\omega^{\prime})
g(k^{\prime},\omega^{\prime})u(k-k^{\prime},\omega-\omega^{\prime})u(k^{\prime},\omega^{\prime})
\end{equation}
For $\epsilon<1$, i.e., for weak disorder, this equation is
iterated to first non--vanishing order in $\epsilon$ ({\it
one-loop} approximation) and averaged over disorder. The disorder
average $\dav{u(k,\omega)u(k^{\prime},\omega^{\prime})}$ can be
calculated using (\ref{eq.UKw}) and (\ref{eq.alphaav}), which
gives
\begin{eqnarray}
\dav{u(k,\omega)u(k^{\prime},\omega^{\prime})}&=& \Lambda^2\pi^2
\left\{\dav{h_{+}(k)h_{-}(k^{\prime})}\delta(\omega-p)\delta(\omega^{\prime}+p)+
\dav{h_{-}(k)h_{+}(k^{\prime})}\delta(\omega+p)\delta(\omega^{\prime}-p)\right\}\nn\\
&=&2\Lambda^2l_{\textit{imp}}\pi^3\delta(k+k^{\prime})\delta(\omega+\omega^{\prime})
\left\{\delta(\omega+p)+\delta(\omega-p)\right\}\nn\\
&\equiv&2\delta(k+k^{\prime})\delta(\omega+\omega^{\prime})D(\omega,k)\,.
\end{eqnarray}
Therefore, we get for $g$ in order $\epsilon^2$
\begin{equation}\label{eq.burgers1loop}
g(k,\omega)=g_0(k,\omega)+\epsilon^2g_0^2(k,\omega)\int\frac{dk^{\prime}\,d\omega^{\prime}}{\Lambda^2(2\pi)^4}
(\omega-\omega^{\prime})\omega^{\prime}\omega(-\omega^{\prime})g_0(k^{\prime},\omega^{\prime})
g_0(k-k^{\prime},\omega-\omega^{\prime})g_0(-k^{\prime},-\omega^{\prime})D(k^{\prime},\omega^{\prime})\,.
\end{equation}

\begin{figure}[htb]
\includegraphics[width=0.6\linewidth]{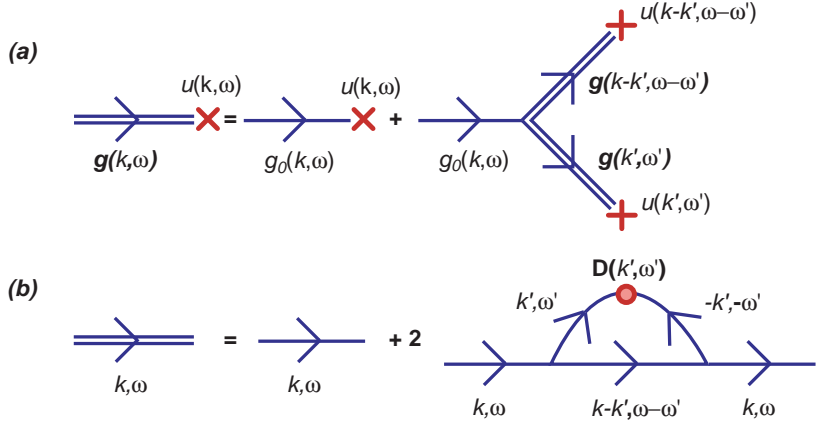}
\caption{Diagrams for (a) eq. (\ref{eq.burgerg}) and (b) eq.
(\ref{eq.burgers1loop}).}\label{fig.burger}
\end{figure}
The diagrams visualizing eqs. (\ref{eq.burgerg}) and
(\ref{eq.burgers1loop}) are depicted in figure \ref{fig.burger}.

For $k=0$ (\ref{eq.burgers1loop}) reduces to
\begin{equation}\label{eq.burgersgk0}
g(0,\omega)=g_0(0,\omega)\left[1+\epsilon^2\frac{p^2l_{\textit{imp}}}{8\pi
t\omega}\int d k^{\prime}\frac{1}{(k^{\prime}/\Lambda)^2+(\pi p^2
t/2)^2}\left\{\frac{p-\omega}{\pi t(\omega-p)^2/2-\imath
k^{\prime}/\Lambda}-\frac{p+\omega}{\pi t(\omega+p)^2/2-\imath
k^{\prime}/\Lambda}\right\}\right]
\end{equation}
\end{widetext}

Because we calculate the correlation length in the thermal regime
(see figure \ref{fig.crossover}) with $t\gtrsim t_u$ the
$k^{\prime}$-integral in (\ref{eq.burgersgk0}) gives the biggest
contribution to $g$ at small $\omega$. In this case the
$k^{\prime}$-integral can be easily calculated which leads to
\begin{equation}
g(0,\omega\ll 1)\approx
g_0(0,\omega)\left(1-\epsilon^2\frac{\Lambda
l_{\textit{imp}}}{2\pi^3p^2 t^3}\right)
\end{equation}
or for the effective Temperature $T_{\textit{eff}}$
\begin{equation}
\frac{1}{T_{\textit{eff}}}\approx\frac{1}{T}\left(1-\epsilon^2\frac{\Lambda
 l_{\textit{imp}}}{2\pi^3p^2 t^3
}\right)=\frac{1}{T}\left(1-\frac{1}{2}\left(\frac{t_u}{\pi p^2
t}\right)^3\right)
\end{equation}
which yields for the correlation length
 \begin{equation}
\xi^{-1}\approx
\frac{\pi}{2}f(T)t\left[1+\frac{1}{2}\left(\frac{t_u}{\pi p^2
t}\right)^3\right]\Lambda\,,
\end{equation}
as written in the text. For high temperatures $t\gg t_u$ we
recover the linear $t$ dependency of the free case.

A related calculation for directed polymers and interface growth
can be found in [\onlinecite{Medina89}]. Note, that in this paper
$x$ plays the role of $\varphi$ and $t$ the role of $x$ in the
above calculation.

\section{Symbol reference}\label{app.sym}

\begin{longtable*}{@{\extracolsep{1cm}}clr}
\caption{\label{tab.sym} List of used quantities}\\
\hline\hline\\[2pt]
symbol& quantity & eq. ref.\\[2pt]
\hline\\[2pt]
\endfirsthead
\multicolumn{3}{c}{\tablename~\thetable: (continued...)}\\[2pt]
\hline\\[2pt]
symbol& quantity & eq. ref.\\[2pt]
\hline\\[2pt]
\endhead
\hline
\endfoot
\hline\hline
\endlastfoot
$a$ & lattice constant & \\[2pt]
$b=e^{-l}$ & rescaling parameter of order 1 & \\[2pt]
$B_i$ & functions used in flow equations & (\ref{eq.B})\\[2pt]
$c$ & elastic constant & (\ref{eq.H0})\\[2pt]
$C(x,\tau)$, $C_f$, $C_b$ & phase correlation functions & (\ref{eq.phaseCF}), (\ref{eq.phaseCFsplit})\\[2pt]
$d$ & spatial dimension & \\[2pt]
$f(T)$ & condensate density & (\ref{eq.qham})\\[2pt]
${\cal F}$ & free energy & (\ref{eq.F})\\[2pt]
$g$ & dimensionless electron-phonon coupling constant & (\ref{eq.qham})\\[2pt]
$g_i$ & functions used in $B_i$ & (\ref{eq.B})\\[2pt]
$g_q$ & electron-phonon coupling constants & (\ref{eq.orderparameter})\\[2pt]
$h_i$ & integers (cf. strong pinning) & (\ref{eq.hi})\\[2pt]
$\hat{\cal H}$, $\hat{\cal H}_0$ & Hamiltonian (complete and free) & (\ref{eq.qham})\\[2pt]
$k$, $k_n$ & wave vectors & \\[2pt]
$k_F$ & Fermi wave vector & \\[2pt]
$K$, $K(l)$, $K_0$ & dimensionless parameter for quantum fluctuations & (\ref{eq.K}), (\ref{eq.dK/dl})\\[2pt]
$K_u$, $K_u^{\ast}$ & $K$-values defining the separatrix/fixed point of the disorder unpinning transition & (\ref{eq.Ku})\\[2pt]
$K_w$, $K_w^{\ast}$ & $K$-values defining the separatrix/fixed point of the lattice unpinning transition &  (\ref{eq.Kw})\\[2pt]
$L$ & system length & (\ref{eq.qham})\\[2pt]
$L_{\textit{FL}}$ & Fukuyama-Lee length& (\ref{eq.FL})\\[2pt]
$l_{\textit{imp}}$ & mean impurity distance & (\ref{eq.UUMW})\\[2pt]
${\cal L}_0$ & free (gaussian) part of the Lagrangian & (\ref{eq.Sn}))\\[2pt]
$n_i$ & integers (cf. strong pinning) & (\ref{eq.strong.pinning.condition2})\\[2pt]
$N_{\textit{imp}}$ & number of impurities & (\ref{eq.Hu}) \\[2pt]
$p$ & commensurability used in the density & (\ref{eq.cdwdensity})\\[2pt]
$p(\epsilon_i)$ & probability density function of $\epsilon_i$ & (\ref{eq.sppdf})\\[2pt]
$\hat P$ & momentum operator, conjugate to $\hat\varphi$ & (\ref{eq.qham})\\[2pt]
$Q$ & density wave vector & (\ref{eq.cdwdensity})\\[2pt]
$q$ & commensurability used in the lattice potential & (\ref{eq.Hw})\\[2pt]
${\cal S}$, ${\cal S}_0$ & action (full and gaussian part) & (\ref{eq.S^n/hbar})\\[2pt]
${\cal S}^{(n)}$ & replicated action & (\ref{eq.Sn})\\[2pt]
${\cal S}_{\textit{SF}}, {\cal S}_{\textit{LP}}$ & action for superfluids and lattice potential, respectively& (\ref{eq.Ssf})\\[2pt]
$S$, $S_1$  & density correlation functions & (\ref{eq.densityCF}), (\ref{eq.densityCFcdw})\\[2pt]
$T$ & temperature & \\[2pt]
$T_c^{\textit{MF}}$ & mean-field condensation temperature & \\[2pt]
$t$, $t(l)$, $t_0$ & parameter for thermal fluctuations & (\ref{eq.t})\\[2pt]
$t_u=1/(\Lambda L_{\textit{FL}})$ & crossover temperature from classical disordered to thermal regime & \\[2pt]
$t_K$ & temperature separating the thermal and disordered regime & \\[2pt]
$U(x)$ & disorder potential & (\ref{eq.Hu})\\[2pt]
$U_i$ & impurity potential & (\ref{eq.Hu})\\[2pt]
$U_{\textit{imp}}$ & mean impurity potential & (\ref{eq.UUMW})\\[2pt]
$u$, $u(l)$, $u_0$ & dimensionless parameter for disorder fluctuations & (\ref{eq.u})\\[2pt]
$v_F$ & Fermi velocity & \\[2pt]
$v$ & phason velocity & (\ref{eq.qham})\\[2pt]
$V_c(x)$ & Coulomb potential & (\ref{eq.HC})\\[2pt]
$W$ & lattice potential strength & (\ref{eq.Hw})\\[2pt]
$w$, $w(l)$, $w_0$ & dimensionless parameter for lattice potential strength & (\ref{eq.w})\\[2pt]
$x_i$ & impurity positions & (\ref{eq.Hu})\\[2pt]
$z$ & dimensionless distance in $\tau$-x-space & \\[2pt]
\hline\\[2pt]
$\alpha$ & parameter used in the strong pinning limit & (\ref{eq.spcf})\\[2pt]
$\beta$ & inverse temperature & \\[2pt]
$\gamma$ & parameter for KT flow equations & (\ref{eq.KTgamma})\\[2pt]
$\Delta$ & order parameter & (\ref{eq.orderparameter})\\[2pt]
$\epsilon_i$ & deviation from mean impurity distance & (\ref{eq.hi})\\[2pt]
$\zeta$ & transverse width of the quasi one-dimensional system & \\[2pt]
$\eta$, $\tilde\eta$ & $=B_0(p^2K_u^{\ast},\infty)$, $=-B_2(q^2/K_w^{\ast},\infty)$, respectively & \\[2pt]
$\lambda$ & density wave length & \\[2pt]
$\lambda_T$ & de Broglie wave length & (\ref{eq.lambdaT})\\[2pt]
$\Lambda$ & momentum cutoff & \\[2pt]
$\xi$, $\xi_u$, $\xi_w$, etc. & correlation lengths & table \ref{tab.cl}\\[2pt]
$\rho(x)$, $\rho_{\textit{SF}}(x)$ & charge/spin or superfluid density & (\ref{eq.cdwdensity}), (\ref{eq.sfdensity})\\[2pt]
$\rho_0$ & mean density & (\ref{eq.cdwdensity})\\[2pt]
$\rho_1$ & density amplitude for harmonic part of $\rho(x)$ & (\ref{eq.cdwdensity})\\[2pt]
$\sigma$ & forward scattering amplitude & (\ref{eq.S^n/hbar})\\[2pt]
$\tau$ & imaginary time coordinate & \\[2pt]
$\varphi$ & phase variable & (\ref{eq.orderparameter})\\[2pt]
$\chi$ & parameter for KT flow equations & (\ref{eq.KTchi})\\[2pt]
$\Upsilon$ & auxiliary function & (\ref{eq.Y})\\[2pt]
$\omega_n$ & Matsubara frequencies & \\[2pt]
\end{longtable*}


\begin{thebibliography}{10}
\bibitem{Fvar}
The inequality for the variational free energy follows directly
from $1-\frac{1}{x}\leq\ln(x)\leq x-1$ for $x>0$, if one considers
the expression $K(\hat\rho_{\textit{SG}},\hat{\tilde\rho})\equiv
Tr \hat{\tilde\rho}
(\ln(\hat{\tilde\rho})-\ln(\hat\rho_{\textit{SG}})$, where
$\hat\rho$ denotes the {\it density matrix} of the system.
\end{thebibliography}
\end{document}